\documentclass[12pt]{article}
\usepackage{amsfonts,epsfig,latexsym,amscd,amsmath,theorem,mathrsfs,color}
\usepackage{algorithm}
\usepackage{algpseudocode}
\usepackage[normalem]{ulem}
\usepackage[dvipsnames]{xcolor}

\graphicspath{{./images/}}
\DeclareMathOperator{\grad}{grad}
\DeclareMathOperator{\diag}{diag}

\newcommand{\notilde}{}

\newcommand{\ket}[1]{\left|#1\right>}

\newcommand{\R}{{\mathbb{R}}}
\newcommand{\Z}{{\mathbb{Z}}}
\newcommand{\N}{{\mathbb{N}}}
\newcommand{\C}{{\mathbb{C}}}
\newcommand{\I}{{\mathbb{I}}}
\newcommand{\DD}{{\mathscr{D}}}

\newcommand{\beq}{\begin{equation}}
\newcommand{\eeq}{\end{equation}}
\newcommand{\bea}{\begin{eqnarray}}
\newcommand{\eea}{\end{eqnarray}}
\newcommand{\ben}{\begin{eqnarray*}}
\newcommand{\een}{\end{eqnarray*}}
\newcommand{\bem}{\begin{enumerate}}
\newcommand{\eem}{\end{enumerate}}

\newcommand{\ra}{\rightarrow}

\newcommand{\cd}{\partial}
\newcommand{\wt}{\widetilde}
\newcommand{\wh}{\widehat}

\def \d{\mathrm{d}}

\newcommand{\ip}[1]{\langle #1 \rangle}

\renewcommand{\star}{*}

\newcommand{\vol}{{\rm vol}}

\newcommand{\ol}{\overline}

\newcommand{\tr}{{\rm tr}\, }

\newcommand{\eps}{\varepsilon}

\theoremstyle{plain}

{\theorembodyfont{\rmfamily}

}
\newcommand{\news}{\setcounter{equation}{0}}

\newcommand{\im}{\mbox{Im}}
\newcommand{\re}{\mbox{Re}}

\newcommand{\figref}{figure \ref}

\begin{document}

\title{Vortex lattices and critical fields in anisotropic superconductors}
\author{
Martin Speight\thanks{E-mail: {\tt j.m.speight@leeds.ac.uk}}\\
School of Mathematics, University of Leeds\\
Leeds LS2 9JT, England \\ \\
Thomas Winyard\thanks{E-mail: {\tt twinyard@ed.ac.uk} (corresponding author)}\\
Maxwell Institute of Mathematical Sciences and School of Mathematics, \\
University of Edinburgh, Edinburgh, \\
EH9 3FD, United Kingdom
}

\maketitle

\begin{abstract}
A method is developed to compute minimal energy vortex lattices in a general Ginzburg-Landau model of  a superconductor subjected to an applied magnetic field. The model may have any number of components and may be spatially anisotropic. The novelty of this method is that it makes no assumptions about the orientation of the vortex lines or the period vectors of the lattice's unit cell: these are all determined dynamically. Methods to compute the first and second critical magnetic fields, $H_{c_1}$ and $H_{c_2}$, in this class of models are also developed.

These methods are applied to a simple anisotropic single-component model, and to an anisotropic two-component model of strong current theoretical interest (a so-called $s+id$ model). It is found, in both cases, that at low applied field the vortex lines can tilt very significantly away from the direction of the applied field (by as much as $40^\circ$ for the single-component and $30^\circ$  for the $s+id$ model). The optimal lattice in the $s+id$ model is qualitatively very different from the conventional triangular Abrikosov lattice, exhibiting a phase transition from a system of Skyrmion chains when the external field is orthogonal to the basal plane to a deformed Abrikosov lattice when applied in the basal plane.  
\end{abstract}

\section{Introduction}
Topological solitons often appear in nature in periodic structures called lattices.  Due to the nonlinear nature of the governing PDEs these periodic solutions must be found numerically.  It is often assumed that the soliton lattice has a particularly nice symmetry group, e.g. the triangular Abrikosov lattice \cite{abrikosov1957magnetic}, but even in spatially homogeneous and isotropic models, soliton lattices often turn out to have much less than maximal symmetry \cite{speight2014crystal, speight2023symmetries,harlandleaskspeight2023}.
In condensed matter contexts, the underlying system is often strongly anisotropic, so there is even less justification in assuming that the
optimal soliton lattice will have high symmetry.  We present a numerical method that allows a general lattice to be found,  minimizing the energy functional with respect to the geometry of a unit cell as well as the fields defined in the cell.  While this approach is general for any model supporting topological solitons, we will focus on the example of the Ginzburg-Landau model of vortices in superconductors under an applied magnetic field, where the extra parameter introduced by the applied field provides some extra challenges.

The response of the interior or bulk of a superconductor when subjected to a constant external magnetic field $H$ has been a key question in physics since it was considered by Peierls and London \cite{peierls1936magnetic, london1936theorie}.  It was observed that as the field strength $|H|$ increased, materials would transition from a superconducting state where the magnetic field was completely expelled, coined the Mei{\ss}ner state,  to a normal metal where the magnetic field penetrated evenly across the material.  To model this transition
 the effective Ginzburg-Landau (GL) model was proposed,  coupling a spatially dependent complex order parameter $\psi$ to the local magnetic field $B$ in the interior of the superconductor.  This effective model was later directly derived from microscopic models at low temperature \cite{gor1959microscopic}.  The model was shown \cite{abrikosov1957magnetic,abrikosov} to exhibit three distinct states separated by two critical values of external field strength $H_{c_1}$ and $H_{c_2}$:
\begin{itemize}
\item  For $|H| < H_{c_1}$ we get the Mei{\ss}ner state or homogeneous superconducting state (MS), where $\psi(x) = u$, $B = 0$ and $u \in \mathbb{C}$ is a constant. 
\item For $H_{c_1} < |H| < H_{c_2}$ we get the mixed (or vortex) state (VS) where $\psi$ and $B$ are inhomogeneous but total magnetic flux through the superconductor is quantized.
\item For $|H| > H_{c_2}$ we get the homogeneous normal state (NS) where $\psi(x) = 0$, $B = H$ and the material acts as a normal metal.
\end{itemize}

It is the mixed state that we will focus on in this paper. The values of $H_{c_1}$ and $H_{c_2}$ are determined by the parameters of the model and it is possible to fix these so that $H_{c_2} \leq H_{c_1}$ and there is no mixed state. Such a superconductor is said to be of type I, while those with $H_{c_2}>H_{c_1}$ are of type II.

\begin{figure}
\begin{center}
\includegraphics[width = 0.9\linewidth]{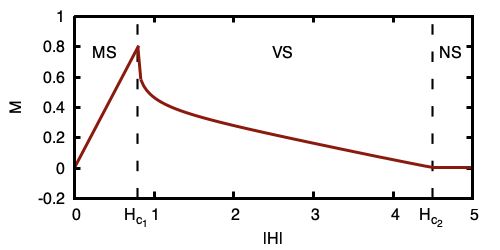}
\end{center}
\caption{The magnetisation $M = |H| - \ip{|B|}$ (difference between strength of applied field and spatial average of the magnetic field) of a standard single-component Ginzburg-Landau model, where the vortex state (VS) is a standard triangular Abrikosov lattice, separating the Mei{\ss}ner state (MS) from the normal state (NS).  This graph was found using the method outlined in section 3, for an isotropic single component superconductor with potential term given by \eqref{eq:Fp} with $\kappa = 3$. }
\label{fig:magCurve}
\end{figure}

If we consider a cross-section perpendicular to $H$, then the mixed or vortex state exhibits topological solitons in the form of vortices.  These vortices are the cross-section of magnetic flux tubes aligned with the direction of $H$ and are topologically preserved.  In a type II superconductor vortices appear in a triangular lattice  \cite{abrikosov1952, abrikosov1957magnetic} coined the Abrikosov lattice. These discrete objects each contribute $2 \pi$ to the total internal magnetic field and as $|H|$ increases in strength the density of the lattice increases. 

The above discussion means that type II superconductors initially expel magnetic field in the Mei{\ss}ner state. Then as $|H|$ increases the magnetic field penetrates parallel to the applied field at discrete points (vortices). The distribution of these discrete points becomes denser until they start to merge.  Finally the magnetic response becomes homogeneous and the material acts as a normal metal.  This process can be seen in \figref{fig:magCurve}.

The above picture, well understood for decades, derives from the simplest realization of GL theory, in which the system is assumed to be spatially isotropic.  In reality, the crystal structure of superconducting materials often introduces significant anisotropies into the Ginzburg-Landau model and, in this case, there is no reason to expect that a symmetric vortex lattice, such as the Abrikosov lattice, will accurately describe the mixed state.  As we will see, even the assumption that the vortex flux tubes are parallel to the applied magnetic field is, in general, ill founded.  In this paper we will present a systematic numerical approach to find the optimal vortex lattice in this case. The novelty of this method is that it makes no assumptions {\it a priori} about the orientation of the vortex lines, or the geometry of the lattice's unit cell. These properties are allowed to vary and are determined dynamically by demanding that the lattice should minimize the system's total Gibbs free energy per unit volume.

Non-symmetric vortex lattices are of particular interest in unconventional superconductors \cite{sigrist1991phenomenological,stewart2017unconventional}, where electrons form Cooper pairs through multiple mechanisms. This modifies the GL model to have multiple, often coupled, order parameters. The couplings between the gradients of these order parameters give vortex solutions, and thus their lattices, interesting new 
properties \cite{winyard2019hierarchies,speight2019chiral}. Examples of such materials include $UTe_2$ \cite{muluneh,cast_of_thousands}, $Mg B_2$\cite{nagamatsu2001superconductivity}, $UPt_3$\cite{joynt2002superconducting} and iron based superconductors\cite{kamihara2008iron}. Our new method was motivated by, and hence naturally lends itself to, probing the lattices of these unconventional materials.

The paper will first present the most general anisotropic Ginzburg-Landau model, then describe a general method for finding the optimal vortex lattice for a given external field $H$ in this general context.  After checking that our method replicates the standard results of Abrikosov under the assumption of spatial isotropy,  we will then discuss how $H_{c_1}$ changes under anisotropy and how to calculate $H_{c_1}$ and $H_{c_2}$. Finally, we will apply the method to two examples, an anisotropic single component model and a multicomponent $s+id$ model.

\section{General Anisotropic GL Model}
The most general anisotropic multi-component Ginzburg-Landau (GL) model has the Gibbs free energy functional,	
\beq
G[\psi_\alpha,A] = \int_{\Omega}\left\{ \frac{1}{2} Q^{\alpha\beta}_{ij} \overline{D_i \psi_\alpha} D_j \psi_\beta + \frac{1}{2} \left| B - H \right|^2 + F_p(\psi_\alpha)\right\},
\label{eq:G}
\eeq
where $D = d - iA$ is the covariant derivative associated with the $U(1)$ gauge field or 1-form $A$.  The local magnetic field is then given by the gauge invariant 2-form $B = dA$.  As we are interested in modelling an infinite superconductor, our space is initially $\Omega = \mathbb{R}^3$.  We will present the method in generality for an $n$-component model, where the $n$ complex fields or order parameters are written $\psi_\alpha =  \rho_\alpha e^{i\varphi_\alpha}$, $\rho_\alpha \in \mathbb{R}_{\geq 0}$ and $\varphi_\alpha \in [0, 2\pi)$ represent the different superconducting bands. Note that Greek indices $\alpha \in [1,n]$ will always enumerate components of the $n$ order parameters and Latin indices $i = 1,2,3$ indicate spatial components, while summation over repeated indices is implied for both. 

$F_p$  collects together the potential terms, which due to gauge invariance, depend only on the condensate magnitudes $\rho_\alpha$ and the phase differences between the condensates $\varphi_{\alpha\beta} := \varphi_\alpha - \varphi_\beta$.  We will also always assume that $F_p$ is bounded below. The energy functional we are using is the Gibbs free energy and thus includes the parameter $H$, interpreted as the externally applied magnetic field which is a constant 2-form. So the energy functional penalizes {\em deviation} of the local magnetic field $B$ from the applied field $H$.

The anisotropy of the model is given by the constant anisotropy matrices $Q^{\alpha\beta}$, which must satisfy the minimal condition $Q^{\alpha\beta}_{ij} = \overline{Q}^{\beta\alpha}_{ji}$ to ensure that the energy is real.  The form of these matrices can, in principle, be derived from the Fermi surfaces of the material under consideration.  Explicit examples will be given later.  We also require that the energy density defined by the anisotropy matrices is positive definite. A convenient way to formulate both the reality and positivity conditions is to collect the complex numbers $Q^{\alpha\beta}_{ij}$ into a single $3n\times 3n$ matrix ${\mathsf Q}_{(\alpha,i)(\beta,j)}$ whose row and column indices range over the set $\{1,2,\ldots,n\}\times\{1,2,3\}$. Then ${\mathsf Q}$ must be Hermitian and positive definite.

The standard isotropic Ginzburg-Landau model can be obtained by simply setting $Q^{\alpha\beta}_{ij} = \delta_{\alpha\beta}\, \delta_{ij}$. If we then choose $n=1$ and set the potential to be,
\beq
F_p(\psi) = \alpha |\psi|^2 + \frac{\beta}{2}|\psi|^4,
\label{eq:Fp}
\eeq
then we have the original (single component) GL model.  Note that this potential only admits superconducting ($|\psi|>0$) solutions for $\alpha < 0$. We will choose to normalize the fields from here on such that the minima of $F_p$ occur for $|\psi|^2 = 1$, hence we introduce a single parameter $\kappa$ (the Ginzburg-Landau parameter) such that $-\alpha = \beta = \kappa^2/2$. Then  $\kappa <1/\sqrt{2}$ gives a type I model (no vortex lattice) while $\kappa >1/\sqrt{2}$ produces a type II model (Abrikosov lattice). 

We are interested in stationary configurations that take the form of local minima of $G$. These satisfy the (bulk) Ginzburg-Landau equations which are obtained by variation of $G$ with respect to the fields $(\psi, A)$,
\begin{align}
Q^{\alpha\beta}_{ij} D_i D_j \psi_\beta &= 2 \frac{\partial F_p}{\partial \overline{\psi}_\alpha},\\
\partial_i ( \partial_j A_i - \partial_i A_j) &= J_i,
\label{eq:eom}
\end{align}
where the total supercurrent is defined as,
\beq
J_i := \im (Q^{\alpha\beta}_{ij} \overline{\psi}_\alpha D_j \psi_\beta). 
\label{eq:current}
\eeq
It is important to note that the parameter $H$ does not appear in the bulk equations of motion. This is not surprising as the only non-constant energy term it appears in is $-\ip{H,\d A}_{L^2}$ which is a topological invariant (it is constant under all variations of $A$ of compact support since $H$ is coclosed).  So $H$ does not affect whether configurations are solutions of the bulk equations of motion. It does affect whether such a solution has minimal energy.

The Ginzburg-Landau equations above always admit two trivial solutions; (NS) the homogeneous normal state $(\psi^{NS}_\alpha,B) = (0,H)$,  and (MS) the homogeneous Mei{\ss}ner state $(\psi^{MS}_\alpha,B) = (u_\alpha,0)$,  where $u_\alpha$ is a constant determined by the form of $F_p$.  Note that we assume that $F_p$ is normalised with respect to the normal state so that $G[\psi^{NS},A^{NS}] = 0$.  We will also define,
\beq
\hat{F}_p(\psi_\alpha) = F_p(\psi_\alpha) - F_p(u_\alpha),
\eeq
which is normalised with respect to the Mei{\ss}ner state so that $\hat{G}[\psi^{MS},A^{MS}] = 0$ where,
\beq
\hat{G}[\psi,A] = G[\psi,A] - \frac{1}{2}\int_\Omega |H|^2 - \int_\Omega F_p(u_\alpha).
\eeq

For $|H|=0$ the Mei{\ss}ner state is the minimal energy solution, while for $|H| \rightarrow \infty$ the normal state is the minimal energy solution. As described in the introduction we are interested in the transition between the MS and NS state where inhomogeneous solutions have minimal energy in the form of topological solitons called vortices. 

These inhomogeneous solutions have only been found numerically and take the form of vortex strings that are translation invariant along the string. Note that the energy functional \eqref{eq:G}, while anisotropic, is translation invariant, so by the Principle of Symmetric Criticality, it is consistent to seek solutions which are invariant under translations in any given fixed direction. So, choose and fix some vector $v_3$ and assume that $\psi_\alpha$ and $A$ are invariant under translation in the direction of $v_3$. Then the energy functional $G$ is infinite, but defines a finite energy per unit length $\ip{G}$, a functional on fields $\psi_\alpha$, $A_i$ defined on $P$, the plane orthogonal to $v_3$. To have finite energy the fields should satisfy the following boundary condition on $P$,
\beq
\rho_\alpha \rightarrow u_\alpha, \quad D \psi_\alpha \rightarrow 0, \quad B \rightarrow H,
\eeq
as $X\ra\infty$, where $X=(X_1,X_2)$ are Cartesian coordinates on $P$. 
Hence $\psi^\infty := \lim_{|X|\rightarrow \infty} \psi(X)$ takes values on a circle of radius $|u|=|(u_1,u_2,\ldots,u_n)|$ in $\C^n$, the $U(1)$ gauge orbit of a vacuum value $u$.  So we have a continuous map $\psi_\infty : S^1_\infty \rightarrow S^1_{u}$ where $S^1_\infty$ is the circle at spatial infinity in $P$ and $S^1_u$ is the orbit of $u$, and the degree, or winding number, of this map is an integer-valued topological invariant of the fields, $N$ (topological because it cannot change under any smooth deformation of the fields preserving finite energy).  The integer $N$ will also correspond to the number of vortices in the system. Then by the boundary condition $D\psi_\alpha \rightarrow 0$ and Stokes's theorem,
\begin{align}
\int_P B_3 &= 2 \pi N,&
\int_P B_1 =  \int_P B_2  &= 0,
\end{align}
where $B_3$ is the magnetic field out of the plane. Hence,  the magnetic flux through $P$ is quantized.  It is important to reiterate that $N$ is topologically conserved in our model, as it is impossible to continuously deform a configuration in one homotopy class into a different class.  However, in a true physical system, materials have finite boundaries where vortices can be created and destroyed. We will not consider this process and focus on the bulk of the superconductor. 

It has been claimed that single component anisotropic models can be reduced to isotropic form by a suitable rescaling of spatial coordinates \cite{klemm1980lower}, but this is mistaken on account of the differing transformation properties of the gradient and magnetic terms in the free energy. Nevertheless, {\em approximate} reduction to an isotropic model is a popular and influential technique in the literature \cite{blatter1992scaling}, particularly for the computation of critical applied fields
\cite{wang2008critical}. As we will see, while the scaling approach of Blatter {\it et al} is reliable for describing states in which the dynamical and applied fields are aligned, it can fail badly outside this context. So the scaling approach predicts $H_{c_2}$ well, but fails to correctly predict $H_{c_1}$.

By contrast, it is well understood that for multicomponent systems ($n>1$) anisotropies cause the length scales of the system to couple, leading to qualitatively different solutions \cite{silaev2018non,winyard2019skyrmion},  caused by non-monotonic inter-vortex interactions.

\section{Finding lattices}\news
Given a choice of the parameter $H$ such that $|H| \geq H_{c_1}$  we seek the minimal energy vortex lattice. This is the global minimizer of the energy per volume $G/|\Omega|$ among fields that are translation invariant in a given direction and doubly periodic orthogonal to this direction up to gauge. Importantly, we must minimize $G/|\Omega|$ not just over the fields for a fixed choice of translation direction and orthogonal period lattice, but also with respect to the choice of direction and lattice. In particular, in an anisotropic model, we are not justified in assuming that the direction of
translation symmetry is parallel to $H$: we must allow it to vary.

\begin{figure}
\begin{center}
\includegraphics[scale = 0.5]{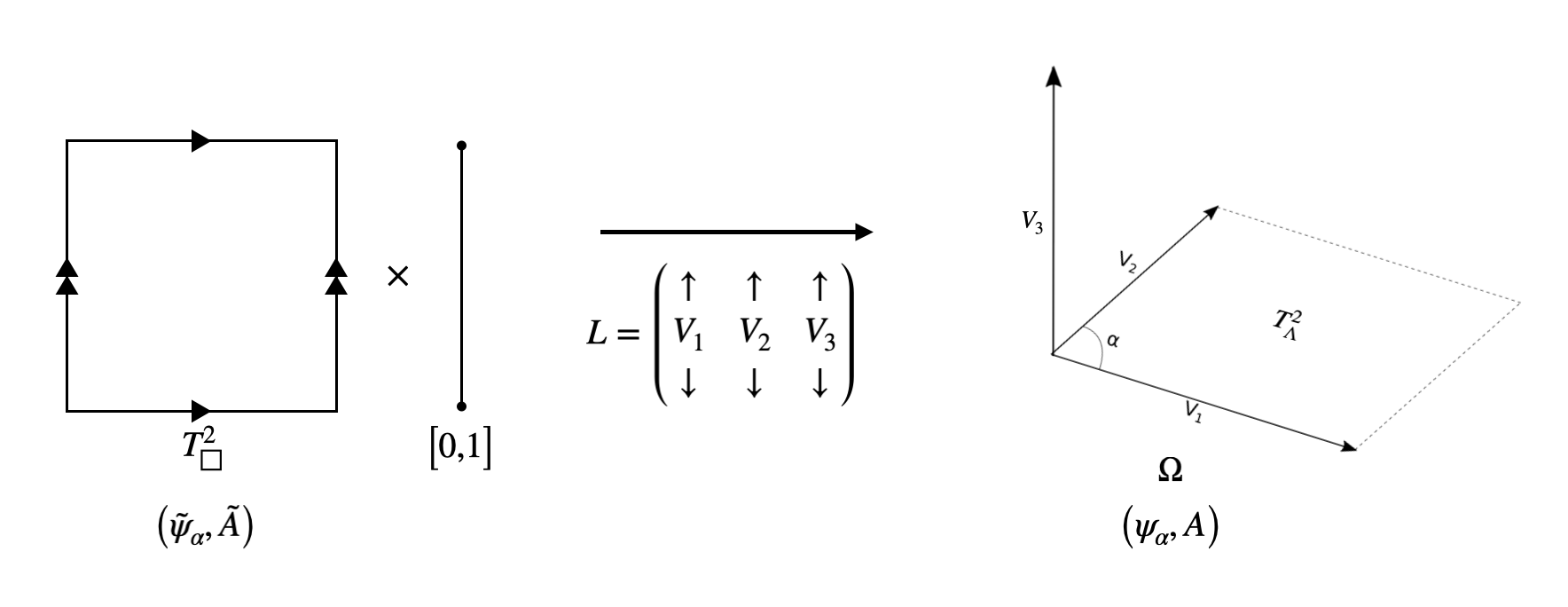}
\end{center}
\caption{The matrix $L$ maps between the fields used in simulation $(\tilde{\psi}_\alpha, \tilde{A})$ defined on a square torus of unit area $T_\square^2$ and the physical fields $(\psi_\alpha, A)$ defined over the physical lattice unit cell $T_\Lambda^2$. }
\label{fig:basis}
\end{figure}

To this end we choose an oriented basis $[v_1, v_2, v_3]$ for $\mathbb{R}^3$, giving the coordinates $X_i$ such that,
\beq
x = L X = X_1 v_1 + X_2 v_2 + X_3 v_3,
\eeq
where $L$ is the matrix whose columns are the chosen basis.  We define a unit cell $\Omega$ as the parallelepiped spanned by $[v_1, v_2, v_3]$, with cell coordinates $X_i$ and volume $|\Omega| = \det L$.  
We will impose translation symmetry in the direction $v_3$ and, without loss of generality, assume that $v_1\cdot v_3=v_2\cdot v_3=0$ (any configuration translation invariant along $v_3$ and doubly periodic in a plane $P'$ not orthogonal to $v_3$ is also doubly periodic in the plane $P$ orthogonal to $v_3$). It is convenient to take 
\beq
v_3=\frac{v_1\times v_2}{|v_1\times v_2|^2}
\eeq
so that the frame is automatically positively oriented, and the unit cell has volume $1$. 

As functions of the coordinates $(X_1,X_2,X_3)$, all fields are independent of $X_3$. In order to allow non-zero magnetic flux per unit area through $P$ (the plane orthogonal to $v_3$) we allow them to be periodic only up to gauge with respect to the lattice spanned by $\{v_1,v_2\}$. In fact, it suffices to impose
\begin{align}
\label{eq:BC1}
\notilde\psi_\alpha(X_1 + 1, X_2) &= \notilde\psi_\alpha(X_1, X_2)e^{i 2\pi N X_2},\\
\notilde\psi_\alpha(X_1, X_2 + 1) &= \notilde\psi_\alpha(X_1, X_2),\\
\notilde{A}_1(X_1 +1, X_2) &= \notilde{A}_1(X_1, X_2),\\
\notilde{A}_2(X_1 +1, X_2) &= \notilde{A}_2(X_1, X_2) + 2\pi N,\\
\notilde{A}_3(X_1 +1, X_2) &= \notilde{A}_3(X_1, X_2),\\
\notilde{A}(X_1, X_2 + 1) &= \notilde{A}(X_1, X_2)
\label{eq:BCfinal}
\end{align}
where $A=A_1\d X_1+A_2\d X_2+ A_3\d X_3$. 
Hence, all gauge invariant quantities $\rho_\alpha$, $\varphi_{\alpha\beta}$, $J_i$ and $B$ are doubly periodic as required.  Note that we have fixed some of our gauge freedom with this choice of boundary conditions. It is also important to note that one must include all components of the gauge field $A$ when the model is anisotropic (including $A_3(X_1,X_2)$), something often neglected in earlier studies of the GL equations. Stokes's Theorem and the above boundary conditions imply that
\beq
\int_{[0,1]^2} \notilde{B}_3 \;  dX_1 \, dX_2 = 2\pi N, \quad \; \int_{[0,1]^2} \notilde{B}_1 \;  dX_1 \, dX_2 =  \int_{[0,1]^2} \notilde{B}_2 \;  dX_1 \, dX_2 = 0,
\eeq
where
\beq
\notilde{B} = \notilde{B}_1 dX_2 \wedge dX_3 + \notilde{B}_2 dX_3 \wedge dX_1 + \notilde{B}_3 dX_1 \wedge dX_2,
\eeq
with
\beq
\notilde{B}_1 = \frac{\partial \notilde{A}_3}{\partial X_2}, \quad \notilde{B}_2 = - \frac{\partial \notilde{A}_3}{\partial X_1}, \quad \notilde{B}_3 = \frac{\partial \notilde{A}_2}{\partial X_1} - \frac{\partial \notilde{A}_1}{\partial X_2}.
\eeq
As we will see, the magnetic field considered as a {\em vector field} rather than a
2-form is simply $B^{\rm vec}=B_1v_1+B_2v_2+B_3v_3$.  This is not entirely obvious, as the basis is not orthonormal, so both the Hodge isomorphism from two-forms to one-forms, and the isomorphism from one-forms to vector fields are nontrivial.
Clearly $N$ determines the number of magnetic flux quanta per unit cell.
 
We can now rewrite the Gibbs free energy in \eqref{eq:G} using the new coordinate system over a single unit cell,
\beq
G = \int_{\Omega} \left\{ \frac{1}{2} M_{ki} Q^{\alpha\beta}_{ij} M^T_{jl} \overline{D_{X_k} \psi_\alpha} D_{X_l} \psi_\beta + \frac{1}{2} \left| B - H \right|^2 + F_P(\psi) \right\} \vol_\Omega,
\eeq
where $M = L^{-1}$ and $\vol_\Omega = \det L \,  dX_1 \wedge dX_2 \wedge dX_3=
dX_1 \wedge dX_2 \wedge dX_3$.  We will now simplify the above expression using some of the assumptions we have made. 

Let us first expand the magnetic term,
\begin{align}
\frac{1}{2} \int_\Omega \left| B - H \right|^2 \vol_\Omega &=  \frac{1}{2}\int_\Omega B \wedge \star B - \int_\Omega B \wedge \star H + \frac{1}{2} \left| H \right|^2 \left|\Omega\right|.
\label{eq:mag}
\end{align}
We write the Euclidean metric in the cell coordinate system as $g = F_{ij} dX_i dX_j$ where $F = L^T L$. We require the action of $\star$ on 2-forms in our coordinates, so define,
\beq
\star\left( dX_1 \wedge dX_2 \right) = \alpha_{3i} dX_i, \quad \star\left( dX_2 \wedge dX_3 \right) = \alpha_{1i} dX_i, \quad \star\left( dX_3 \wedge dX_1 \right) = \alpha_{2i} dX_i.
\eeq
The induced inner product on 1-forms is $\left< dX_i, dX_j \right> = \left(F^{-1}\right)_{ij}$ and,  by definition, for any 2-form $\mu$ and 1-form $\lambda$ we have $\left< \star \mu, \lambda \right> \vol_\Omega = \mu \wedge \lambda$. Hence, taking $\mu = dX_1 \wedge dX_2$ and $\lambda = dX_j$ we get,
\begin{align}
\alpha_{3i} \left(F^{-1}\right)_{ij} &= \delta_{j3}.
\end{align}
Rearranging this and repeating the process with $\mu = dX_2 \wedge dX_3$ and $\mu = dX_3 \wedge dX_1$ we see that,
\beq
\alpha_{ij} =  {F_{ij}},
\eeq
whence
\beq
*B=B_iF_{ij}\d X_j,
\eeq
so
\beq
B \wedge \star B = {B_{i} B_{j} F_{ij}} \, dX_1 \wedge dX_2 \wedge dX_3.
\eeq
Further $g(B^{\rm vec},\cd/\cd X_i)=*B(\cd/\cd X_i)=B_jF_{ji}=g(B_j\cd/\cd X_j, \cd/\cd X_i)$, so
\beq
B^{\rm vec}=B_j\frac{\cd\: }{\cd X_j}=B_1v_1+B_2v_2+B_3v_3.
\eeq

For the second term in \eqref{eq:mag} we use Stokes's theorem,
\begin{align}
\int_\Omega B \wedge \star H &= \int_\Omega d\left(A \wedge \star H\right),\\
&= \int_{\partial \Omega} A \wedge \star H.
\end{align}
It will be useful to define $\tilde{H}_j dX_j := \star H = H_i dx_i = H_i L_{ij} dX_j $. We then consider each of the square faces that compose $\partial \Omega$ at $X_i = 0$ or $1$,
\begin{align}
\int_\Omega B \wedge \star H &= \int_{X_1 = 1} \left( A_2 \tilde{H}_3 - A_3 \tilde{H}_2\right) \, dX_2 \wedge dX_3 - \int_{X_1 = 0} \left( A_2 \tilde{H}_3 - A_3 \tilde{H}_2\right) \, dX_2 \wedge dX_3\nonumber\\
&+ \int_{X_2 = 1} \left( A_3 \tilde{H}_1 - A_1 \tilde{H}_3\right) \, dX_3 \wedge dX_1 - \int_{X_2 = 0} \left( A_3 \tilde{H}_1 - A_1 \tilde{H}_3\right) \, dX_3 \wedge dX_1\nonumber\\
&+ \int_{X_3 = 1} \left( A_1 \tilde{H}_2 - A_2 \tilde{H}_1\right) \, dX_1 \wedge dX_2 - \int_{X_3 = 0} \left( A_1 \tilde{H}_2 - A_2 \tilde{H}_1\right) \, dX_1 \wedge dX_2,\\
&= \int_{[0,1]^2} \left( \tilde{H}_3 ( A_2 (1, X_2, X_3) - A_2(0, X_2, X_3) \right) dX_2 \, dX_3,\\
&= 2 \pi N H_i L_{i3},
\end{align}
where we have made use of the boundary conditions on $A$ above. 

If we now recombine the above terms we can write the Gibbs free energy per unit volume to which we seek global minimizers,
\beq
\left< G \right> = \frac{G}{|\Omega |} = \frac{1}{2} M_{ki} P_{ki, lj} M_{lj} + \frac{1}{2} \tr ( L \mathbb{B} L^T) - 2N \pi H_i L_{i3} + \frac{1}{2} |H|^2 + \int_{[0,1]^2} F_p(\psi),
\label{eq:<G>}
\eeq
where we have introduced,
\begin{align}
P_{ki,lj} &= \re \int_{[0,1]^2} Q^{\alpha\beta}_{ij} \overline{D_{X_k}\psi_\alpha } D_{X_l}\psi_\beta \, dX_1\, dX_2, \\
\mathbb{B}_{ij} &= \int_{[0,1]^2} B_i B_j \, dX_1\, dX_2.
\end{align}

Having written $\left< G \right>$ in terms of the cell basis, one can then use any standard numerical scheme to minimise it with respect to the fields $(\psi_\alpha, A)$ subject to a fixed unit cell $L\in SL(3,\R)$. However, we also need to determine a numerical scheme to minimise $\left< G \right>$ with respect to $L \in SL(3,\mathbb{R})$ for a fixed field configuration.  As $L$ only appears in the first 3 terms of \eqref{eq:<G>} (recall $M=L^{-1}$)
we need only consider those terms.  Recall we have assumed, without loss of generality, that
\beq
L = \left( v_1 \quad v_2 \quad \frac{v_1 \times v_2}{|v_1 \times v_2|^2} \right)
\label{eq:L}
\eeq
for some linearly independent pair $v_1,v_2\in \R^3$, so that $\Omega$ has unit volume and the plane $P$ spanned by $v_1,v_2$ is orthogonal to $v_3$. 
Minimising $\left<G\right>$ subject to these constraints is equivalent to minimising $\left<G\right>$ on the codimension 3 algebraic variety $\mathcal{C} \subset GL(3, \mathbb{R}) \subset \mathbb{R}^9$ on which,
\begin{align}
\label{eq:con1}
\det L = 1,\\
\label{eq:con2}
L_{i1}L_{i3} = 0,\\
L_{i2}L_{i3} = 0.
\label{eq:con3}
\end{align}
Note that the first condition is cubic, so is not (as in the two-dimensional analogue of this problem) the level set of a quadratic form \cite{speight2023symmetries,speight2023magnetic}. For this reason we have been unable to
minimize $\ip{G}$ over $\mathcal{C}$ explicitly and have resorted to numerics.  Note also that \eqref{eq:L} gives an explicit parametrization of $\mathcal{C}$ in terms of the local coordinates $(v_1, v_2)$. 

The numerical goal is now, given a fixed configuration $(\psi_\alpha, A)$, to minimize $\left< G \right>$ in \eqref{eq:<G>} over $\mathcal{C}$.  Let $L(t)$ be a curve in $\mathcal{C}$ through $L = L(0)$ with $\dot{L}(0) = \eps$. Then,
\begin{align}
M(t) L(t) &= I_3,\\
\dot{M}(0) L(0) + M(0) \dot{L}(0) &= 0,\\
\dot{M}(0) &= - M(0) \eps M(0),
\end{align}
leading to,
\beq
\left.\frac{d}{dt}\right|_{t=0} \left< G \right> \left( L(t) \right) = \eps_{ik}\left( - M_{qp} P_{qp, lj} M_{li} M_{kj} + L_{ij} \mathbb{B}_{jk} - 2N\pi H_i \delta_{k3} \right).
\eeq
Hence, the gradient of $\left< G \right>:\mathcal{C}\ra\R$ at $L\in\mathcal{C}$ tangent to $\mathcal{C}$ is,
\beq
\left( \grad\left< G \right> \right)_{ik} = P_{\mathcal{C}} \left( - M_{qp} P_{qp, lj} M_{li} M_{kj} + L_{ij} \mathbb{B}_{jk} - 2N\pi H_i \delta_{k3} \right),
\eeq
where $P_{\mathcal{C}}$ denotes orthogonal projection tangent to $\mathcal{C}\subset\R^9$. The projector $P_{\mathcal C}:\R^9\ra T_L\mathcal{C}$ is 
straightforward to construct numerically via a Gramm-Schmidt algorithm, starting from the basis of coordinate basis vectors for $T_L{\mathcal{C}}$ defined by the parametrization \eqref{eq:L}.

Once we have the projector we can minimize $\left< G \right>$ to a given tolerance $g_{tol}$ with respect to the unit cell $L$ (for a  fixed field configuration $(\psi_\alpha, A)$) via a simple gradient descent algorithm.  We evolve $L$ as,
\beq
L \mapsto L - dt \left( \grad\left< G \right> \right)
\eeq
repeating until the absolute values of all components of $\grad\left< G \right>$ are smaller than $g_{tol}$. 

We now turn to finding the fields that minimize $\left< G \right>$ subject to a fixed $L$. Note that any standard numerical method for finding minima of energy functionals for field theories would work here, for example gradient flow. In addition, as $L$ is fixed, finding the minimum of $\left< G \right>$ is equivalent to finding the minimum of $G$ over the unit cell. 

We discretize the cross-section of the unit cell on a regular two-dimensional grid of $N_1 \times N_2$ lattice sites with spacing $h>0$.  We then approximate the 1st and 2nd order spatial derivatives using central 4th order finite difference operators,  which yields a discrete approximation $\left< G \right>_{dis}:S\ra\R$ to the functional in \eqref{eq:<G>}, where the discretized configuration space is the manifold $S = (\C^n \times \mathbb{R}^3)^{N_1\times N_2} \cong \R^{(2n+3)N_1N_2}$.  We then seek local minima of $\left< G \right>_{dis}$ subject to the boundary conditions given in \eqref{eq:BC1} - \eqref{eq:BCfinal}.  We evolve the system using a gradient descent method, namely the arrested Newton flow algorithm (described in detail in \cite{speight2020skyrmions}), solving for the motion of a particle in $S$ under the potential $\left< G \right>_{dis}$, 
\beq
\ddot{\Phi} = -\grad \left<G\right>_{dis}\left( \Phi \right),
\eeq
starting at the initial configuration $\Phi(0)$ and $\dot{\Phi}(0) = 0$ (here $\Phi$ denotes our collective discretized fields, a point in $S$).  Evolving this algorithm will cause the configuration to relax towards a local minimum. At each time step $t \rightarrow t + \delta t$, we check to see if the direction of the force on the particle opposes its velocity. If $\dot\Phi(t)\cdot \grad\ip{G}_{dis}(\Phi(t))>0$, then we set $\dot{\phi}_{t + \delta t} = 0$ and restart the flow.  The flow is terminated once  every component of $\grad \left< G \right>_{dis}(\phi)$ is zero within a given tolerance.

We now have an algorithm to find the optimal vortex lattice,  given $H$ with $H_{c_1} < |H| < H_{c_2}$, an initial unit cell $L_0$ and an initial field configuration $(\psi_\alpha, A)_0$ that satisfies the boundary conditions with topological degree $N$ given in \eqref{eq:BC1}-\eqref{eq:BCfinal}. 

\section{Spatially isotropic systems}\news
	
It is important that our method replicates the standard results of Abrikosov in the case of an isotropic single component type II superconductor. Namely, we consider $n=1$ and $Q_{ij} = \delta_{ij}$ such that 
\beq
P_{ki,lj} = \delta_{ij} \mathbb{P}_{kl}, \quad \mathbb{P}_{kl} = \re \int_{[0,3]^2} \overline{D_{X_k} \psi} D_{X_l} \psi \, dX_1 dX_2.
\eeq
Hence, 
\beq
\left< G \right>(L) = \frac{1}{2} \tr (M M^T \mathbb{P}) + \frac{1}{2} \tr ( L \mathbb{F} L^T) - 2N\pi H_i L_{i3} + \frac{1}{2} |H|^2 + \int V.
\eeq
For general rotation $R \in SO(3)$, $(RL)^{-1} = M R^T$, so
\begin{align}
\left< G \right>(RL) &= \frac{1}{2} \tr ( M R^T R M^T \mathbb{P}) + \frac{1}{2} \tr ( RL\mathbb{F} L^T R^T ) - 2N \pi H_i R_{ik} L_{k3} + \frac{1}{2} |H|^2 + \int V,\\
&= C(L)- 2N \pi H_i R_{ik} L_{k3}=C(L)-2N \pi H\cdot (Rv_3),
\end{align}
where $C(L)$ denotes terms independent of $R$.
Hence, the minimum of $\left< G \right>$ over the $SO(3)$ orbit of a given matrix $L$ occurs when $(RL)_{i3} = k H_i$ for some $k > 0$, when we rotate the cell so that the translation symmetry direction is aligned with $H$. Hence, the minimal energy lattice configuration must have vortex lines parallel to the applied field $H$.  Note, that this argument holds for an isotropic model with any number of components ($n \geq 1$) as then $Q^{\alpha\beta}_{ij} = \delta_{ij} q_{\alpha\beta}$ where $q$ is a hermitian matrix. The above argument then proceeds unchanged but with,
\beq
\mathbb{P}_{kl} = \re \int_{[0,3]^2} q_{\alpha\beta} \overline{D_{X_k} \psi_\alpha} D_{X_l} \psi_\beta \, d X_1\, dX_2.
\eeq

Now that we know that the vortex line will align with the external field,  if we consider all the energy terms dependent on $A_3$ it is no longer coupled and the choice $A_3=0$ minimizes them. This means that $B_1 = B_2 = 0$ and hence the internal magnetic field is always parallel to $H$, or $B(x) = b(x) H_i$ where $b(x) \in \mathbb{R}$.  

We have applied the numerical scheme described above to a single component isotropic model, with $F_p$ given in \eqref{eq:Fp} in the type~II regime ($\kappa > 1/\sqrt{2}$), with $H=(0,0,H)$ and $v_1,v_2$ in the $x_1-x_2$ plane. We find that the energetically optimal lattice is triangular, as expected, that is $|v_1|=|v_2|$ and the angle between them is $60^\circ$, and $N=2$. Hence, our scheme reproduced the Abrikosov vortex lattice in this simple case.

\section{Finding $H_{c_1}$}\news
The critical value $H_{c_1}$ represents the smallest strength of external field $|H|$ such that there exists a vortex state with lower Gibbs free energy $G$ than the Mei{\ss}ner state ($\psi_\alpha = u_\alpha$, $B = 0$). It is important to note that, in an anisotropic model, $H_{c_1}$ depends on the direction of $H$. 
Moreover, when constructing the minimal energy vortex state for a given $H$, we should {\em not} assume that the vortex is translation invariant in the $H$ direction: just as in the computation of optimal lattices, in general there may be vortices with lower energy that have $v_3$ non-parallel to $H$. 

For a given external field $H$ and degree $N$, we construct the minimal energy degree $N$ vortex as follows. We choose a unit vector $v_3$ and an orthonormal basis $\{v_1,v_2\}$ for the plane $P$ orthogonal to $v_3$, then
assign to any collection of fields $(\psi,A)$ translation invariant in the $v_3$ direction and  decaying to $u$ with winding $N$ on the boundary of $P$, its Gibbs free energy per unit length, normalized so that the Mei{\ss}ner state has energy $0$, that is,
\beq
\hat{G}[\psi,A]=\int_{P}\left\{\frac12 (D\psi)^\dagger QD\psi+\frac12|B|^2+F_p(\psi)\right\}-2\pi N H\cdot v_3.
\eeq
It is important to realize that this quantity depends on $v_3$ not only through the explicit dependence of the final term, but also through the dependence of the $Q$ matrices on the orientation of $P$.
For fixed $[v_1,v_2,v_3]$ we minimize $\hat{G}$ with respect to the fields
(by arrested Newton Flow, for example). This produces a function $\hat{G}_{min}:SO(3)\ra \R$, mapping the frame $[v_1,v_2,v_3]$ to the Gibbs free energy of the minimal $N$-vortex aligned with the $v_3$ axis, which we minimize by gradient flow. (In fact, $\hat{G}_{min}$ descends to a function on $S^2=SO(3)/SO(2)$, since the energy actually only depends on $v_3$, but by treating it as a function on $SO(3)$, we may repurpose the gradient flow algorithm used to find optimal lattices  to solve this problem too: we simply project the flow to the submanifold of ${\mathcal{C}}$ on which $\{v_1,v_2\}$ are orthonormal.) So, for a fixed $H$ and $N$, we have a minimal $N$-vortex with Gibbs free energy $\hat{G}(H,N)$. For $|H|$ small, $\hat{G}(H,N)>0$, while for $|H|$ sufficiently large $\hat{G}(H,N)<0$. The degree $N$ lower critical field for a given applied field direction $\hat{H}=H/|H|$ is the smallest $H_0$ for which $\hat{G}(H_0\hat{H},N)=0$. Let us denote this
$H_{c_1}^N(\hat{H})$. The lower critical field for the direction $\hat{H}$ is then
\beq
H_{c_1}(\hat{H})=\inf_{N\in\Z}H_{c_1}^N(\hat{H}).
\eeq
In practice, we compute $H_{c_1}^N(\hat{H})$ for a small selection of degrees (typically $N=1$ and $2$ only), and assume $H_{c_1}$ is the minimum of these.

Associated to $H_{c_1}(\hat{H})$ there is an optimal degree $N$ vortex solution, translation invariant in some direction $v_3(\hat{H})$. It is important to realize that, in general, there is no reason why $v_3$ should equal $\hat{H}$: the optimal vortex at the threshold for flux penetration may have vortex lines (and magnetic flux) which are {\em not} aligned with the applied magnetic field, if the underlying system is anisotropic. We will see that this observation holds even in the case of single component models.

\section{Finding $H_{c_2}$}\news

It is clear that, for any constant applied field $H$, the normal state
$\psi=0$, $B=H$ is a solution of the field equations \eqref{eq:eom}, and hence a critical point of $G$. It is not necessarily a {\em stable} critical point of $G$ (a local minimum) however. To test stability of the normal state, we consider the {\em second} variation of $G$.

 Let $\psi_t$, $A_t$ be a smooth variation of the fields with $\psi_0=0$ and $\d A_0=H$ (so we are varying about the normal state). Let $\eps:=\cd_t\psi_t|_{t=0}$ and
$\eta:=\cd_t A_t|_{t=0}$, the infinitesimal generators of the variation.
The normal state is linearly stable if $d^2 G[\psi_t,A_t]/dt^2|_{t=0}\geq 0$ for all such variations. A routine calculation yields
\beq
\frac{d^2\:}{dt^2}\bigg|_{t=0}G[\psi_t,A_t]
=\int_{\Omega}\ol{\eps_\alpha}\left\{-Q^{\alpha\beta}_{ij}D_iD_j\eps_\beta+M_{\alpha\beta}\eps_\beta\right\}+\int_\Omega|\d\eta|^2
\eeq
where
\beq\label{mdef}
M_{\alpha\beta}:=2\frac{\cd^2 F_p}{\cd\ol\psi_\alpha\cd\psi_\beta}\bigg|_{\psi=0},
\eeq
whence it is clear that the normal state is linearly stable if and only if the self-adjoint linear operator 
\beq
(\wh{O}\eps)_\alpha:=-Q^{\alpha\beta}_{ij}D_iD_j\eps_\beta+M_{\alpha\beta}\eps_\beta
\eeq
has non-negative spectrum. This section presents a general numerical method to address this linear stability criterion, and hence extract $H_{c_2}$, the upper critical field of the system. 

First we choose and fix a unit vector $\hat{H}$ and consider applied fields in this direction, so $H=|H|\hat{H}$. We then rotate our coordinate system so that the 3rd coordinate points along the $H$ direction. This amounts to choosing $R\in SO(3)$ with 3rd column $\hat{H}$ and defining new coordinates $(X_1,X_2,X_3)$ such that $x=RX$. This transforms the $Q$ matrices
\beq
Q^{\alpha\beta}\mapsto \mathcal{Q}^{\alpha\beta}=R^TQ^{\alpha\beta}R.
\eeq
In this coordinate system, the gauge field producing $B=|H|\d X_3$ may be chosen to be
\beq
A= \frac{|H|}{2}(-X_2\d X_1+X_1\d X_2).
\eeq
It is convenient to rescale the coordinates,
\beq
Y_i:=\sqrt{\frac{|H|}{2}}X_i
\eeq
so that the covariant derivatives are 
\beq
D_i=\sqrt{\frac{|H|}{2}}\DD_i,\qquad
\DD_1=\frac{\cd\:}{\cd Y_1}+iY_2,\quad
\DD_2=\frac{\cd\:}{\cd Y_2}-iY_1,\quad
\DD_3=\frac{\cd\:}{\cd Y_3}.
\eeq
The operator whose spectrum we seek is now
\beq
(\wh{O}\eps)_\alpha=-\frac{|H|}{2}\mathcal{Q}^{\alpha\beta}_{ij}\DD_i\DD_j\eps_\beta+M_{\alpha\beta}\eps_\beta.
\eeq
Note that all the $|H|$ dependence of this operator is now explicit. Denote
 by $\lambda_0$ the lowest eigenvalue of $\wh{O}$. Assuming the system's temperature is below $T_c$, the matrix $M$ has at least one negative eigenvalue, while the operator $-\mathcal{Q}^{\alpha\beta}_{ij}\DD_i\DD_j$ is manifestly positive. Hence, for $|H|=0$, $\lambda_0<0$ (and the normal state is unstable), while for $|H|$ sufficiently large,
      $\lambda_0>0$ (and the normal state is stable). $H_{c_2}$ is,
       by definition, the value of $|H|$ at which the sign 
       of $\lambda_0$ changes. 

It remains to compute the least eigenvalue of $\wh{O}$. We first note that
$[\wh{O},\I_n\otimes i\DD_3]=0$, so we may seek simultaneous eigenstates of $\wh{O}$ and $\I_n\otimes i\DD_3$. Hence, we may assume our eigenstate takes the form
\beq\label{kk}
\eps=\phi(Y_1,Y_2)e^{ikY_3}
\eeq
for some $k\in\R$. All previous studies of $H_{c_2}$ that we are aware of assume that the ground state has $k=0$. It transpires, however, that in general this assumption is not valid: one can certainly construct systems whose ground state has $k\neq0$, as we will see shortly, so we will not make this assumption here. To proceed further, we define operators
\beq
a:=\frac{i}{2}(\DD_i+i\DD_2),\qquad
a^\dagger=\frac{i}{2}(\DD_1-i\DD_2),\qquad
\nu:=a^\dagger a,
\eeq
and note that these satisfy the harmonic oscillator algebra
\beq
[\nu,a^\dagger]=a^\dagger,\qquad
[\nu,a]=-a,\qquad
[a,a^\dagger]=1,
\eeq
so $a,a^\dagger$ are ``ladder" operators for the ``number" operator $\nu$. 
Eliminating $\DD_1,\DD_2$ in favour of $a,a^\dagger$ we see that, on the
$k$-eigenspace of $\I_n\otimes i\DD_3$, our operator takes the form 
\begin{equation}
\wh{O}_k = \frac{|H|}{2}\left\{L_1(a^\dagger)^2 + L_1^\dagger a^2 + L_2 (a^\dagger a + a a^\dagger)+L_3+k(L_4a^\dagger+L_4^\dagger a)+k^2L_5\right\} + M
\end{equation}
where we have defined the $n\times n$ complex matrices,
\begin{align}
L_1^{\alpha\beta} &= \mathcal{Q}^{\alpha\beta}_{11} - \mathcal{Q}^{\alpha\beta}_{22} + i\left( \mathcal{Q}^{\alpha\beta}_{12} + \mathcal{Q}^{\alpha\beta}_{21}\right)\\
L_2^{\alpha\beta} &= \mathcal{Q}^{\alpha\beta}_{11} + \mathcal{Q}^{\alpha\beta}_{22}\\
L_3^{\alpha\beta} &= i\left( \mathcal{Q}^{\alpha\beta}_{12} - \mathcal{Q}^{\alpha\beta}_{21}\right)\\
L_4^{\alpha\beta} &=-\left(\mathcal{Q}^{\alpha\beta}_{13}+\mathcal{Q}^{\alpha\beta}_{31}\right)+i\left(\mathcal{Q}^{\alpha\beta}_{23}+\mathcal{Q}^{\alpha\beta}_{32}\right)\\
L_5^{\alpha\beta} &= \mathcal{Q}^{\alpha\beta}_{33},
\end{align}
and $M$ was defined in \eqref{mdef}.

We now define the functions 
\begin{align}
\left|0\right> &:= e^{-(Y_1^2 + Y_2^2)/2}\\
\left|p\right> &:= \frac{1}{\sqrt{p!}}(a^\dagger)^p\left|0\right>
\end{align}
and note that $a\left|0\right> = 0$, $a\ket{p}=\sqrt{p}\ket{p-1}$ and
$a^\dagger\ket{p}=\sqrt{p+1}\ket{p+1}$. We then seek eigenfunctions of $\wh{O}_k$ of the form
\beq
\phi_\alpha = \sum_{p=0}^{\infty}c_p^\alpha \left|p\right>.
\eeq
That is, our function space is $\Phi:=\C^n\otimes S$ where $S=\bigoplus_{p\in\N}
\ket{p}$ is the space of ``particle" states spanned by $\ket{p}$. It is useful to introduce the parity operator
\beq
P_S:S\ra S, \qquad P_S\ket{p}=(-1)^p\ket{p},
\eeq
and define $P:=\I_n\otimes P_S:\Phi\ra\Phi$.
Note that $P_S$ anti-commutes with both $a$ and $a^\dagger$, and so for all
$\phi\in\Phi$,
\beq
\wh{O}_k P\phi=P\wh{O}_{-k}\phi.
\eeq
It follows immediately that the spectrum of $\wh{O}$ is symmetric under $k\mapsto -k$, since if $\wh{O}_k\phi=\lambda\phi$ then 
$\wh{O}_{-k}P\phi=\lambda P\phi$. Hence the lowest eigenvalue $\lambda_*(k)$ of $\wh{O}_k$ attains a local extremum at $k=0$, and it suffices to compute $\lambda_*(k)$ for $k\geq 0$. Since $L_5$ is a positive hermitian matrix, it is clear that $\lambda_*(k)$ grows unbounded above quadratically as $|k|\ra\infty$, so $\lambda_*(k)$ attains a minimum at some $k_0\in[0,\infty)$, and $\lambda_0=\lambda_*(k_0)$. As, just observed, it seems to be a universal assumption in the literature that $\lambda_*(k)$ attains a minimum at $k=0$, but for multicomponent ansisotropic systems, this is not necessarily true. 

To compute $\lambda_*(k)$ numerically, we truncate the state space $S$ to finite dimension $S_m=\bigoplus_{p=0}^m\ket{p}$ so that the ladder operators are approximated by the finite matrices
\beq
a_m = \left( \begin{array}{cccccc} 
0 & \sqrt{1} & 0 & 0 & \dots & 0\\
0 & 0 & \sqrt{2} & 0 & \dots & 0\\
0 & 0 & 0 & \sqrt{3} & \dots & 0\\
\vdots & \vdots & \vdots & \vdots &  & \vdots\\
0 & 0 & 0 & 0 & \dots & \sqrt{m} \\
0 & 0 & 0 & 0 & \dots & 0 \end{array}\right),\qquad
a_m^\dagger = a_m^T.
\eeq
This produces a $n(m+1)\times n(m+1)$ matrix $O_k^{(m)}$ approximant to $\wh{O}_k$,
\bea
O_k^{(m)}&=&\frac{|H|}{2}\left\{L_1\otimes(a_m^\dagger)^2 + L_1^\dagger \otimes a_m^2 + L_2\otimes (a_m^\dagger a_m + a_m a_m^\dagger)+L_3\otimes \I_{m+1}\right.\nonumber \\&&\qquad\left.+k(L_4\otimes a_m^\dagger+L_4^\dagger\otimes a_m)+k^2L_5\otimes\I_{m+1}\right\} + M\otimes\I_{m+1},
\eea
whose lowest eigenvalue $\lambda_m$ can be 
computed by any standard linear algebra package. We start with $m=2$ and keep doubling $m$ until $|\lambda_{m}-\lambda_{m/2}|$ is less than some desired tolerance, at which point we accept the approximation $\lambda_*(k)=\lambda_m$.  The results presented below had tolerance $10^{-9}$, which typically required a state space with $m=64$.  

This algorithm computes the lowest eigenvalue $\lambda_*(k)$ for a given value of $k\in[0,\infty)$. One must then use a one-dimensional search method to find the minimum value attained by this quantity as $k$ varies. This minimum value is $\lambda_0$, the lowest eigenvalue of $\wh{O}$ for the applied field $H=|H|\hat{H}$, in the chosen direction $\hat{H}$. In general, one must then vary $|H|$ to find where $\lambda_0(|H|)$ crosses from negative to positive, this value being $H_{c_2}$ for the direction $\hat{H}$.

This final search problem (with respect to the parameter $|H|$) can be avoided if, as is the case in most models of phenomenological interest,
$\psi=0$ is a local maximum of $F_p$. In this case, the matrix $M$ is negative definite. Let $e_1,e_2,\ldots,e_n$ be a unitary basis of eigenvectors of $M$ corresponding to the eigenvalues $-\mu_1^2,-\mu_2^2,\ldots,-\mu_n^2$. Then we may perform a linear transformation of our fields by defining $\wt\psi_\alpha$ so that
\beq
\psi=\sum_{\alpha=1}^n\frac{\wt\psi_\alpha}{\mu_\alpha}e_\alpha,
\eeq
whence, with respect to the new fields,
\beq
F_p=-\frac12\wt\psi^\dagger\wt\psi+O(|\wt\psi|^3),
\eeq
and the corresponding $M$-matrix is $\wt{M}=-\I_n$.
The Gibbs energy takes the same form \eqref{eq:G} in the new fields, but with transformed anisotropy matrices
\beq
\wt{Q}_{ij}=D^{-1}U^\dagger Q_{ij}UD^{-1},
\eeq
where $U$ is the unitary matrix whose columns are $e_1,\ldots,e_n$, and 
$D=\diag(\mu_1,\mu_2,\ldots,\mu_n)$. Hence, the normal state $\wt\psi=0, B=H$ is linearly stable if and only if the operator
\beq
\wh{O}=-\frac{|H|}{2}\wt{\mathcal{Q}_{ij}}\DD_i\DD_j-\I_n
\eeq
has positive spectrum, where $\wt{\mathcal{Q}}^{\alpha\beta}=R^T\wt{Q}^{\alpha\beta}R$ are the spatially rotated $Q$-matrices, as before. But the lowest eigenvalue of this operator is $|H|\wt{\lambda}_0-1$ where $\wt{\lambda}_0$ is the lowest eigenvalue of the $|H|$-independent operator
\beq
\wt{O}:=-\wt{\mathcal{Q}}_{ij}\DD_i\DD_j.
\eeq
Hence $H_{c_2}=2/\wt{\lambda}_0$, and we need only find the lowest eigenvalue of the single operator $\wt{O}$. Note that the search over $k$ is still necessary to find $\wt{\lambda}_0$. Note also that $\wt{\lambda}_0$, and hence $H_{c_2}$ in general depend on $\hat{H}$, the direction of the applied field, through the $R$-dependence of the matrices $\wt{\mathcal{Q}}$.
 
\begin{figure}
\begin{center}
\includegraphics[width = \linewidth]{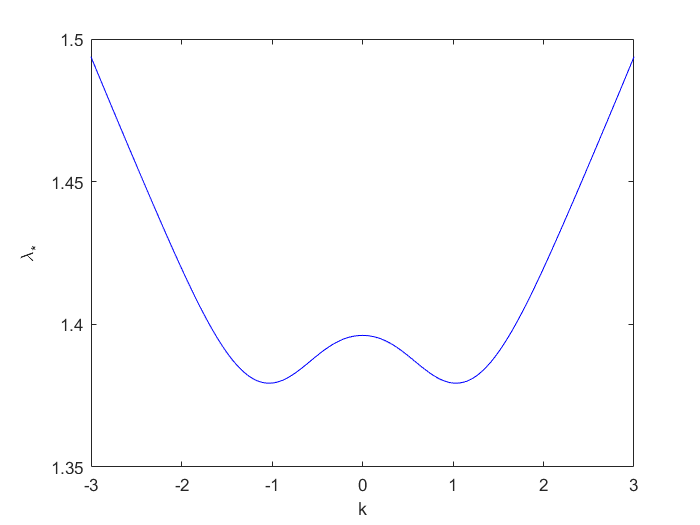}
\end{center}
\caption{The lowest eigenvalue, $\lambda_*$, of the stability operator $\wt{O}$ as a function of $Y_3$-momentum, $k$, for the anisotropic two-component model with $Q$ matrices defined in \eqref{eq:crazyQs}. Note that this eigenvalue attains a local {\em maximum} at $k=0$, not a minimum.}
\label{fig:weird_Hc2}
\end{figure}

This is the algorithm used to compute $H_{c_2}$ for all of the specific models considered in this paper. We end this section by applying it to a simple two-component model devised to illustrate that the ground state may have $k\neq 0$. The model has
\beq\label{eq:crazyQs}
Q^{11}=Q^{22}=\I_3,\qquad
Q^{12}=\left(\begin{array}{ccc}
-0.35&   -0.25 &   0.39 \\
   -0.24&   0.11&   0.38 \\
    0.42 &    0.37&   -0.4
\end{array}\right)
+i\left(\begin{array}{ccc}
0.11&    0.21&    0.27 \\
         0 &  -0.1 &    0.07 \\
    0.18&    0.14&   0.22
\end{array}\right)
\eeq
and $M=-\I_2$. Choosing $\hat{H}=(0,0,1)$, we find that the lowest eigenvalue $\lambda_*(k)$ of $\wt{O}$ acting on the $k$-eigenspace of $\I_2\otimes\DD_3$ has a local {\em maximum} at $k=0$, and attains its minimum at $k=\pm 1.03$. Making the erroneous assumption that the ground state has $k=0$ would lead us to underestimate $H_{c_2}$ by 1.2\% for this model and field orientation.
A graph of $\lambda_*(k)$ is presented in \figref{fig:weird_Hc2}.

The model \eqref{eq:crazyQs} was engineered to have a stability operator whose ground state has $k\neq 0$. It turns out that all the models we will consider subsequently do not exhibit this exotic behaviour. Nonetheless, this is something that must be checked on a case by case basis.

\section{Anisotropic single component GL}\news
We will now make the simplest extension to the type II example  considered in section 4 by introducing spatial anisotropy.  Consider the single component model with anisotropy matrix,
\beq
Q^{11} = \left( \begin{array}{ccc} 1 & 0 & 0 \\ 0 & 1 & 0 \\ 0 & 0 & \lambda_z \end{array}\right).
\label{eq:singleQ}
\eeq
and potential $F_p$ as in \eqref{eq:Fp} with $\kappa = 3$ (strongly type II).  Physically $1/\lambda_z$ gives an effective mass ratio of the electron excitations for different spin directions. This model has an $SO(2)$ symmetry about the $z$-axis but will exhibit markedly different behaviour dependent on the angle that the applied field $H$ makes with the basal ($x$--$y$) plane, which we denote $\theta$ (defined so that $H_3=|H|\cos\theta$). 

\begin{figure}
\begin{center}
\includegraphics[width = \linewidth]{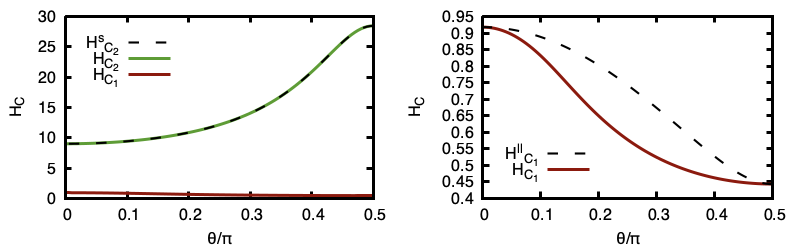}
\end{center}
\caption{A plot of the critical values of external field strength $H_{c_1}$, $H_{c_2}$ (left) and $H_{c_1}$ alone (right) for the single component model given in \eqref{eq:singleQ} with $\kappa = 3$. They are plotted for all orientations of the external field $H = |H|( \cos \phi \sin \theta,  \sin \phi \sin \theta, \cos \theta)$ parametrised by $\theta$. Note that the model has an $SO(2)$ symmetry about the $z$-axis and hence is invariant w.r.t. $\phi$. The dashed curve in the left plot ($H^s_{c_2}$) is the prediction from the rescaling method given in \eqref{eq:rescaling}. The dashed line in the right plot ($H^{\parallel}_{c_1}$) corresponds to the value of $H_{c_1}$ (erroneously) predicted if we assume that the vortex lines must be parallel to $H$.
 }
\label{fig:singleHc}
\end{figure}

Experimentally, the clearest manifestation of spatial anisotropy is in the ratio of
second critical fields $H_{c_2}^z$ and $H_{c_2}^x=H_{c_2}^y$ in the $z$-direction and
 in the basal plane respectively. For many strongly anisotropic materials this ratio ranges between 2 and 3 \cite{klemm1980lower,  kidszun2011critical, tang2017anisotropic, gu2018single}. In our model, $\lambda_z=0.1$ produces a model with
\beq 
\Gamma:= \frac{H^x_{c_2}}{H^z_{c_2}}=3.16.
\eeq
 This choice is also consistent with models in the literature  \cite{klemm1980lower, tang2017anisotropic}, so we fix $\lambda_z=0.1$ for all our simulations.

\begin{figure}
\begin{center}
\includegraphics[width = 0.89\linewidth]{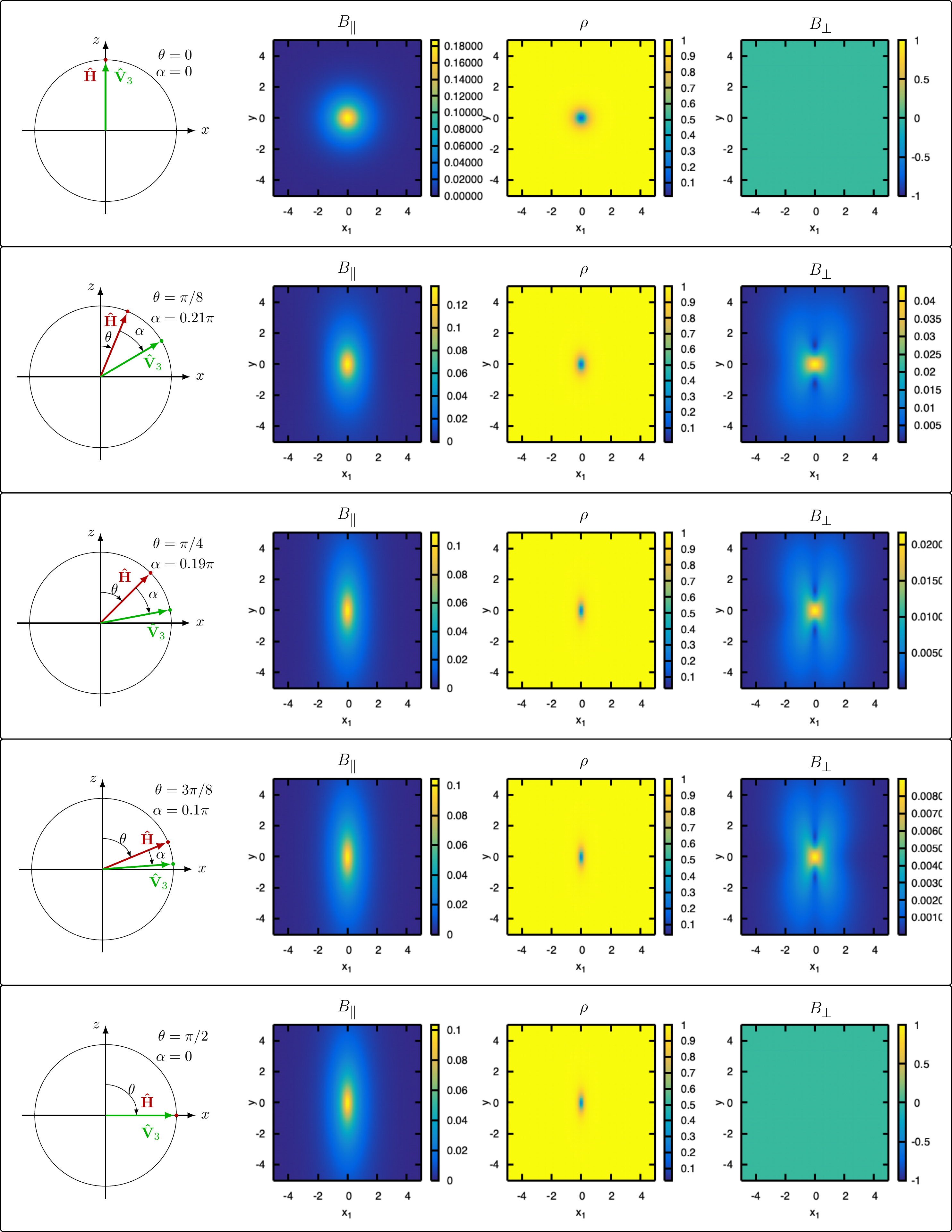}
\end{center}
\caption{Plots of vortices for the single component model in \eqref{eq:singleQ} for $|H| = H_{c_1}$ and $\hat{H} = (\sin \theta, 0, \cos \theta)$, found using the algorithm described in section 5.  The left panel shows the applied field direction $\hat{H}$, the normal to the vortex plane $\hat{v}_3$ and the angle between them $\alpha$. The other panels are a cross-section of the vortex line  in the plane spanned by $\hat{x}_1 = (\cos(\theta + \alpha), 0, -\sin(\theta + \alpha))$ and $\hat{y}=(0,1,0)$, where $B_\parallel = B\cdot v_3,$ $\rho=|\psi|$, and $B_\perp = |B-B_\parallel v_3|$ (orthogonal to the vortex line).}
\label{fig:singleStatic}
\end{figure}

The critical fields for this model are plotted in \figref{fig:singleHc}.  Note that $H_{c_1}$ is a decreasing function of  $\theta\in[0,\pi]$ whereas $H_{c_2}$ monotonically increases.  This is unsurprising as the effective mass is increasing, making the model effectively more type II and hence increasing the region of parameter space for which the vortex state is optimal.  We have also compared the difference in approximating $H_{c_1}$ by assuming the vortex line and applied field are parallel (dashed line) and by solving the full non-linear problem presented in section 5 (solid red line).  As predicted when applied along one of the crystalline axes the two methods agree, however the true $H_{c_1}$ is as low as $77\%$ of the approximation $H^\parallel_{c_1}$.

It is interesting to compare $H_{c_1}(\theta)$ and $H_{c_2}(\theta)$ with the expressions predicted by the scaling method of \cite{blatter1992scaling}. Our single component model has (in their notation) anisotropy parameter $\eps^2=\lambda_z=0.1$. Hence, any ``magnetic quantity" $Q(\theta)$ in the anisotropic model with field
applied at angle $\theta$ is obtained from the same ``magnetic quantity" $\wt{Q}$ for a related 
(but fixed) isotropic model by
\beq
Q(\theta)=\frac{\wt{Q}}{\sqrt{\cos^2\theta+\eps^2\sin^2\theta}}.
\label{eq:rescaling}
\eeq
In deriving this formula, one should note that \cite{blatter1992scaling} uses a different definition of $\theta$, and we are using their scaling formula for a magnetic quantity $Q$ which does not itself depend on the applied field strength -- as $H_{c_1}$ and $H_{c_2}$ tautologically do not.
Note that formula \eqref{eq:rescaling} implies identical $\theta$ dependence for all such magnetic quantities, and is increasing on $[0,\pi]$ if $\eps^2<1$, as in our case. Clearly, this is qualitatively quite wrong for $H_{c_1}(\theta)$, but it reproduces $H_{c_2}(\theta)$ perfectly.

\begin{figure}
\begin{center}
\includegraphics[scale = 0.5]{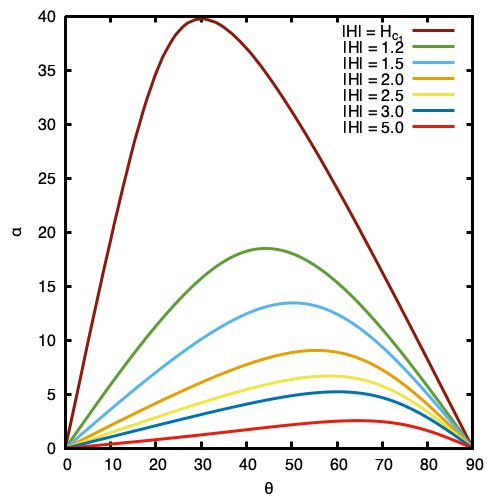}
\end{center}
\caption{A plot of the angle $\alpha$ between the applied field $H$ and the vortex line $\hat{v}_3$ for optimal vortex lattices at various field strengths $|H|$ and orientations $\theta$ such that $H = |H|(\cos \phi \sin \theta, \sin \phi \sin \theta, \cos \theta)$. Note that the model is invariant w.r.t. $\phi$.  The top red curve corresponds to the limiting case $|H(\theta)|\searrow H_{c_1}(\theta)$. In the opposite limit, $|H(\theta)|\nearrow H_{c_2}(\theta)$, $\alpha(\theta)\ra 0$.}
\label{fig:singleAngle}
\end{figure}

The static configurations that were found using the $H_{c_1}$ algorithm are displayed for various applied fields in \figref{fig:singleStatic}.  The left panel shows the angle $\alpha$ between the applied field $H$ and the vortex line $v_3$, this is also plotted as the red curve in \figref{fig:singleAngle}.  This angle is high when the applied field is far from the crystaline axes $\hat{x}$ or $\hat{z}$ and is almost $\pi/4$ at its highest. When the angle $\alpha$ is high we see that the magnetic field twists direction in the plane, as indicated by the right panel in the plots.

\begin{figure}
\begin{center}
\includegraphics[width=0.96\linewidth]{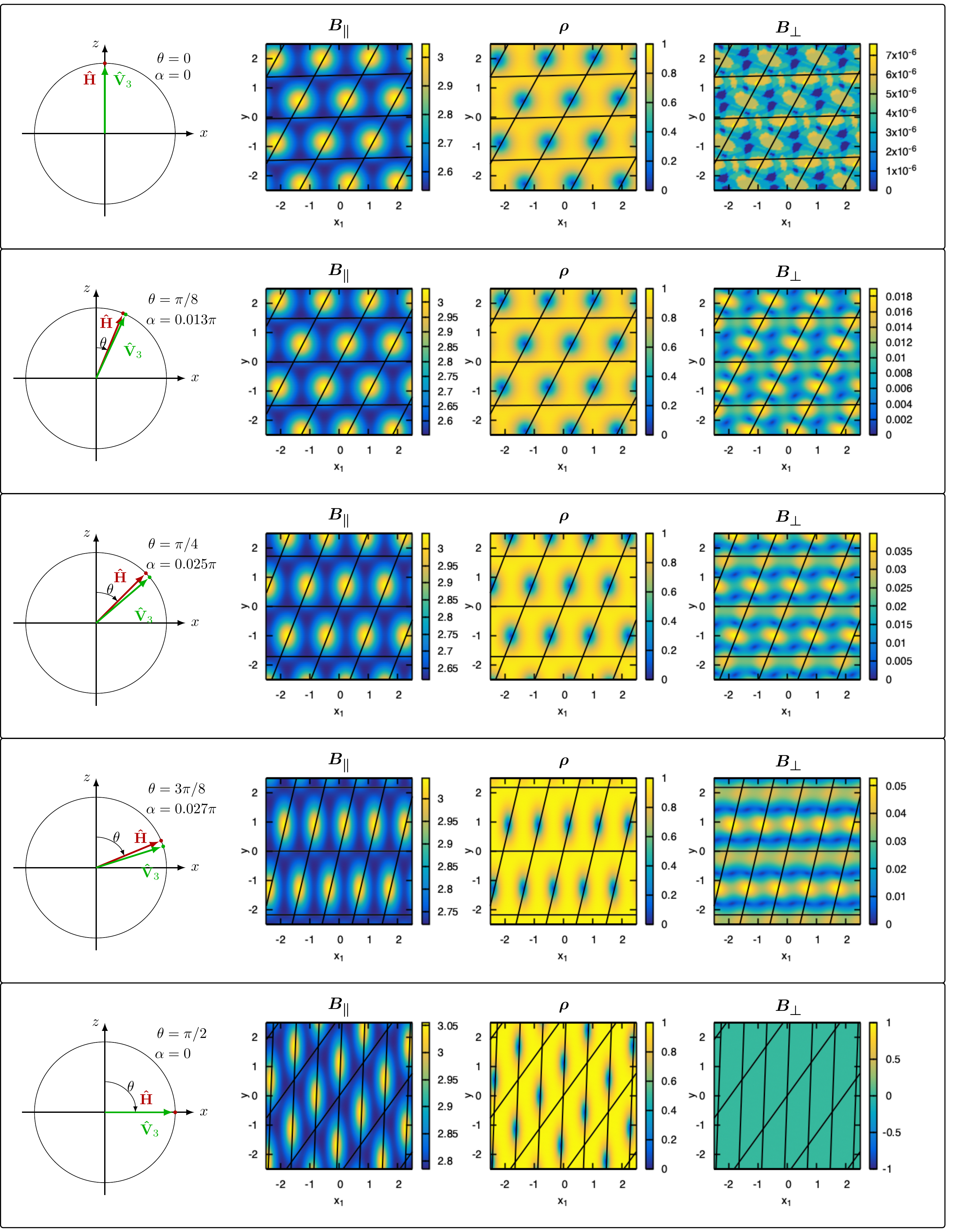}
\label{fig:singleLattice}
\end{center}
\caption{Plots of the vortex lattice minimizers for the single component model in \eqref{eq:singleQ} for $|H| = 3$ and $\hat{H} = (\sin \theta, 0, \cos \theta)$, found using the algorithm in section 3. The notation and labels are the same as \figref{fig:singleStatic}. }
\end{figure}

Finally we have plotted some lattice solutions for various applied field directions $\hat{H}$ for an applied field strength of $|H| = 3$ in \figref{fig:singleLattice}. We can see that for the top row ($\hat{H} = \hat{z}$) we have an exact Abrikosov lattice and for the bottom row ($\hat{H}=\hat{x}$) we have a stretched Abrikosov lattice.   We then continuously deform the lattice as the angle of $\hat{H}$ with $\hat{z}$ changes from $0$ to $\pi/2$. As the lattice deforms the angle between $\hat{H}$ and $\hat{v}_3$ increases, which is shown clearly for many values of $|H|$ in \figref{fig:singleAngle}.  As the applied field is increased the angle becomes shallower and as it decreases the curve approaches the limiting curve for when $|H| = H_{c_1}$. 

\section{Multicomponent $s+id$ example}
\begin{figure}
\begin{center}
\includegraphics[width = 0.49\linewidth]{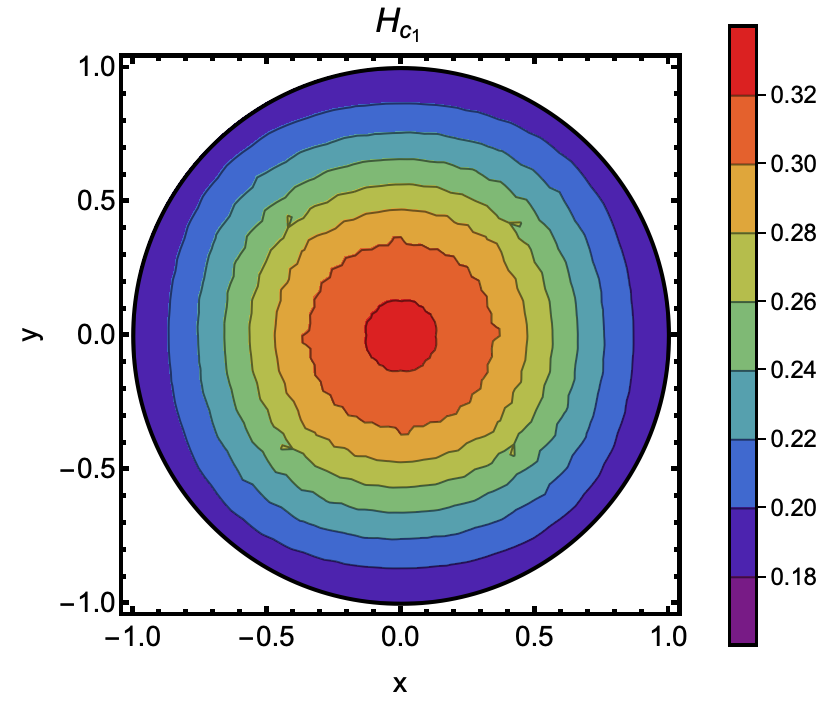}
\includegraphics[width = 0.49\linewidth]{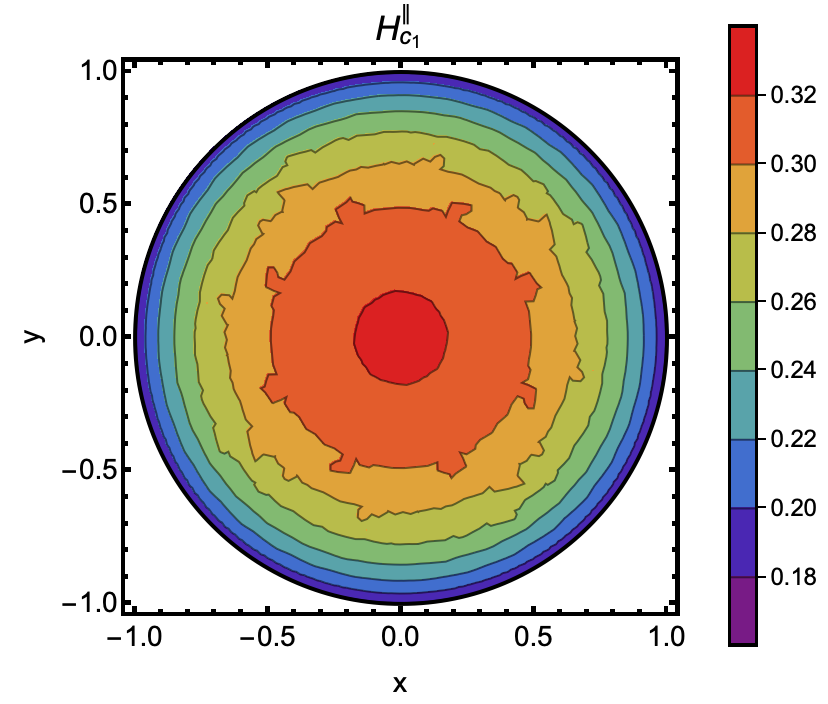}
\includegraphics[width = 0.49\linewidth]{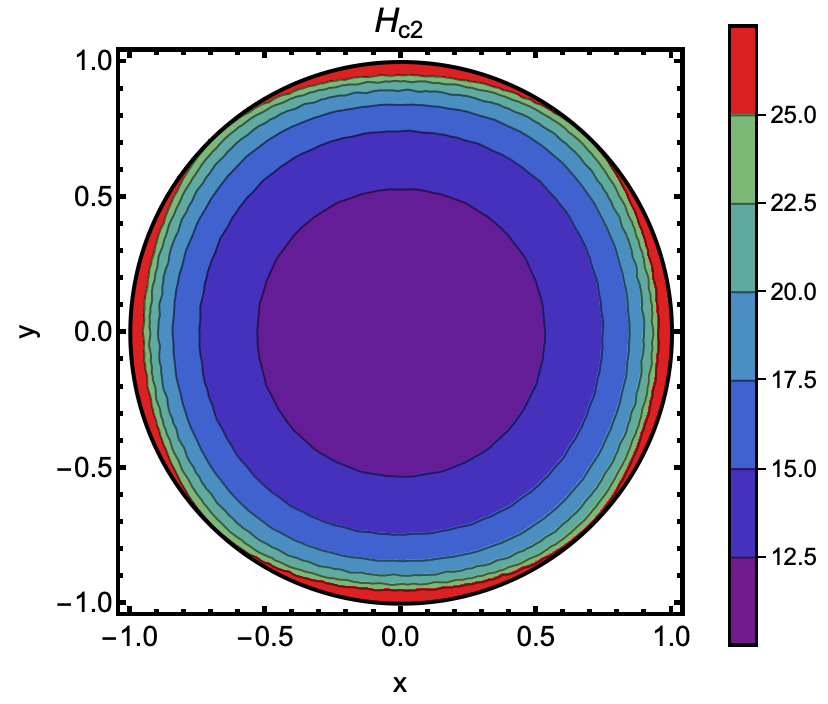}
\end{center}
\caption{The correct values of $H_{c_1}$ (left),  the approximation $H_{c_1}^\parallel$ which is calculated assuming that $v_3$ is parallel to $H$ (right), and $H_{c_2}$ (bottom) for the $s+id$ model given in \eqref{eq:sidAni} and \eqref{eq:sidFp},  calculated using the methods presented in section 5 and 6.  The horizontal and vertical axes correspond to the $x$ and $y$ coordinates of the normalised applied field $\hat{H}$ and the colour the strength of $H_{c_1}$ and $H_{c_2}$. Note that while the scale is the same for $H_{c_1}$ and $H_{c_1}^\parallel$, it differs for $H_{c_2}$ as the parameters give a strongly type II model. Also the value for $H_{c_1}$ ($H_{c_2}$) decreases (increases) from the origin $\hat{H}=\hat{z}$ radially to the equator, leading to a more type II model when $\hat{H}$ is in the basal plane (equator).  }
\label{fig:sidHc}
\end{figure}

We now consider a more complicated example with both anisotropy and multiple components.  In particular, we will consider an $s+id$ model derived in \cite{ren1995ginzburg}.  This is an $n=2$ component model that exhibits a $d_{x^2 - y^2}$ electron pairing symmetry, which is of interest in modelling high $T_c$ superconductors in materials such as $YBCO$ \cite{wollman1993experimental}.  To write this model in our notation we use,
\beq
Q^{11} =  \frac{4}{\sqrt{2}\lambda} \, \mbox{diag}(1, 1, \kappa), \quad \;
Q^{22} = \frac{2}{\sqrt{2}\lambda}\mbox{diag}(1, 1, \kappa),\quad \;
Q^{12} = \frac{2}{\sqrt{2}\lambda}\mbox{diag}(1, -1, 0)
\label{eq:sidAni}
\eeq
with the potential,
\beq
F_p = -\alpha_\alpha |\psi_\alpha|^2 + \frac{\beta_\alpha}{2} |\psi_\alpha|^4 + \gamma_{12} |\psi_1|^2 |\psi_2|^2 + \eta_{12} |\psi_1|^2 |\psi_2|^2 \cos \varphi_{12},
\label{eq:sidFp}
\eeq
Here $\kappa$ is a parameter that has been introduced as the model in \cite{ren1995ginzburg} is 2-dimensional, focussing entirely on solutions in the basal $(x,y)$ plane with applied field always orthogonal to this in the $z$ direction.  For the rest of the paper we will make use of the following parameters,
\begin{align}
\nonumber\alpha_1 &= 1.4,&
\alpha_2 &=1,&
\beta_1 &= \frac{4}{3},&
\beta_2 &= \frac{1}{2},\\
\gamma_{12} &= \frac{8}{3},&
\eta_{12} &= \frac{4}{3},&
\lambda &=4.
\label{eq:sidPot}
\end{align}
To approximate physical values for $\kappa$ we can consider the anisotropy of the crystal unit cell which is orthorhombic \cite{gu2018single} with a cell of,
\beq
a=b= 3.8677 \mbox{\r{A}}, \quad c = 12.2874 \mbox{\r{A}},
\eeq
where \r{A} is Angstroms.  This gives an aspect ratio of $c/a = 3.1769$.   Hence, by applying the relative rescalings we get that $\kappa = (a/c)^2 \approx 0.1$.  The best way to check this naive approximation is physically sensible is to compare $H_{c_2}$ anisotropy with that from experimental data.  If we define,
\beq
\Gamma := \frac{H_{c_2}^{\mbox{basal}}}{H_{c_2}^{z}},
\eeq
where ``basal" refers to the $x$--$y$ plane,
then experiment suggests that $\Gamma \approx 2$ \cite{wang2008critical}.  
For our chosen parameters $\Gamma$ is between 2.5 and 3.5 depending on the choice of basal direction, where $H/|H| = (1,0,0)$ has the highest $H_{c_2}$ and $H/|H| = (1,1,0)$ has the smallest $H_{c_2}$, note that the model has 4-fold symmetry about the $z$-axis.  The different values for $H_{c_2}$ can be seen in \figref{fig:sidHc}. In addition we can see that $H_{c_1}$ follows a similar pattern but reversed, so it is higher when $\hat{H}$ is out of the basal plane and more suppressed when $\hat{H}$ is pointing in the plane.

Note that this $s+id$ model has previously been considered in the basal plane \cite{zhang2020skyrmionic}.  In this paper the authors studied large rectangular systems of vortices and studied the skyrmion chains that formed.  It has also been shown that the coupled length scales of this model \cite{speight2021magnetic} lead to a lot of the unconventional behaviour it exhibits \cite{benfenati2020magnetic}.  Some of these details were also subsequently summarised in \cite{wormald2021topological}.

\begin{figure}
\begin{center}
\includegraphics[scale = 0.5]{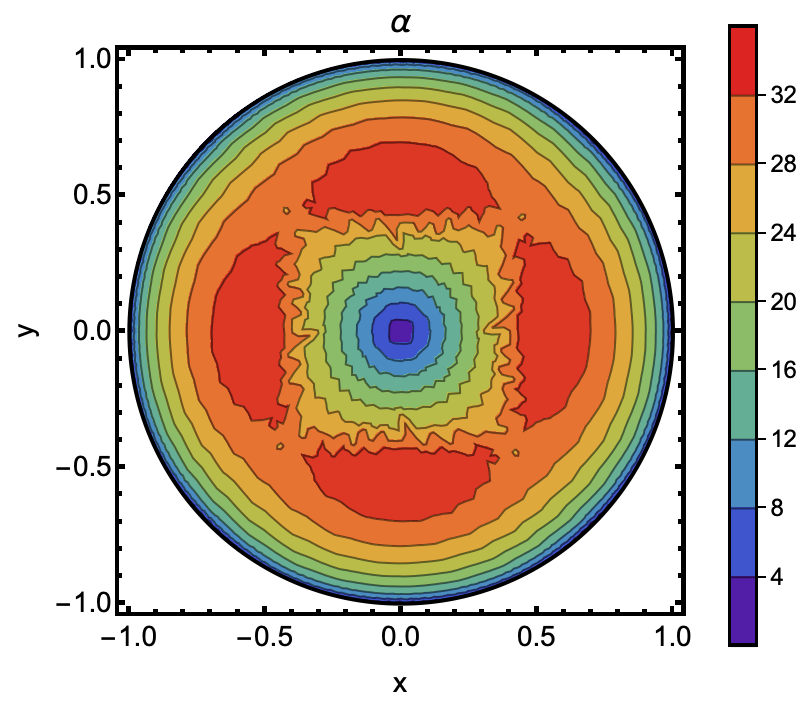}
\end{center}
\caption{A heat plot of the angle $\alpha$ between the applied field $H$ and the vortex line $\hat{v}_3$ in degrees, for the $s+id$ model given in \eqref{eq:sidAni} and \eqref{eq:sidFp} where $|H| = H_{c_1}$.  The horizontal and vertical axes correspond to the $x$ and $y$ coordinates of the normalised applied field $\hat{H}$.  As predicted the angle is zero at the origin $\hat{H}=\hat{z}$ and for $\hat{H}$ in the basal plane (equator). The plot retains the four fold symmetry of the model, and hits its peak along the $x=0$ and $y=0$ axes.}
\label{fig:sidAngle}
\end{figure}

We have also plotted the angle that $\hat{H}$ makes with the vortex line in \figref{fig:sidAngle}, for $|H| = H_{c_1}$. The maximal angle of deviation is close to $30^\circ$, which is a substantial change.  In addition we can also see the deviation of $H_{c_1}$ from the old method where it is assumed $\hat{v}_3$ and $H$ are parallel ($H_{c_1}^{\parallel}$) in \figref{fig:sidHc}. These two plots demonstrate the importance of taking into account the vortex line disinclination to the applied field $H$. 

\begin{figure}
\begin{center}
\includegraphics[width = \linewidth]{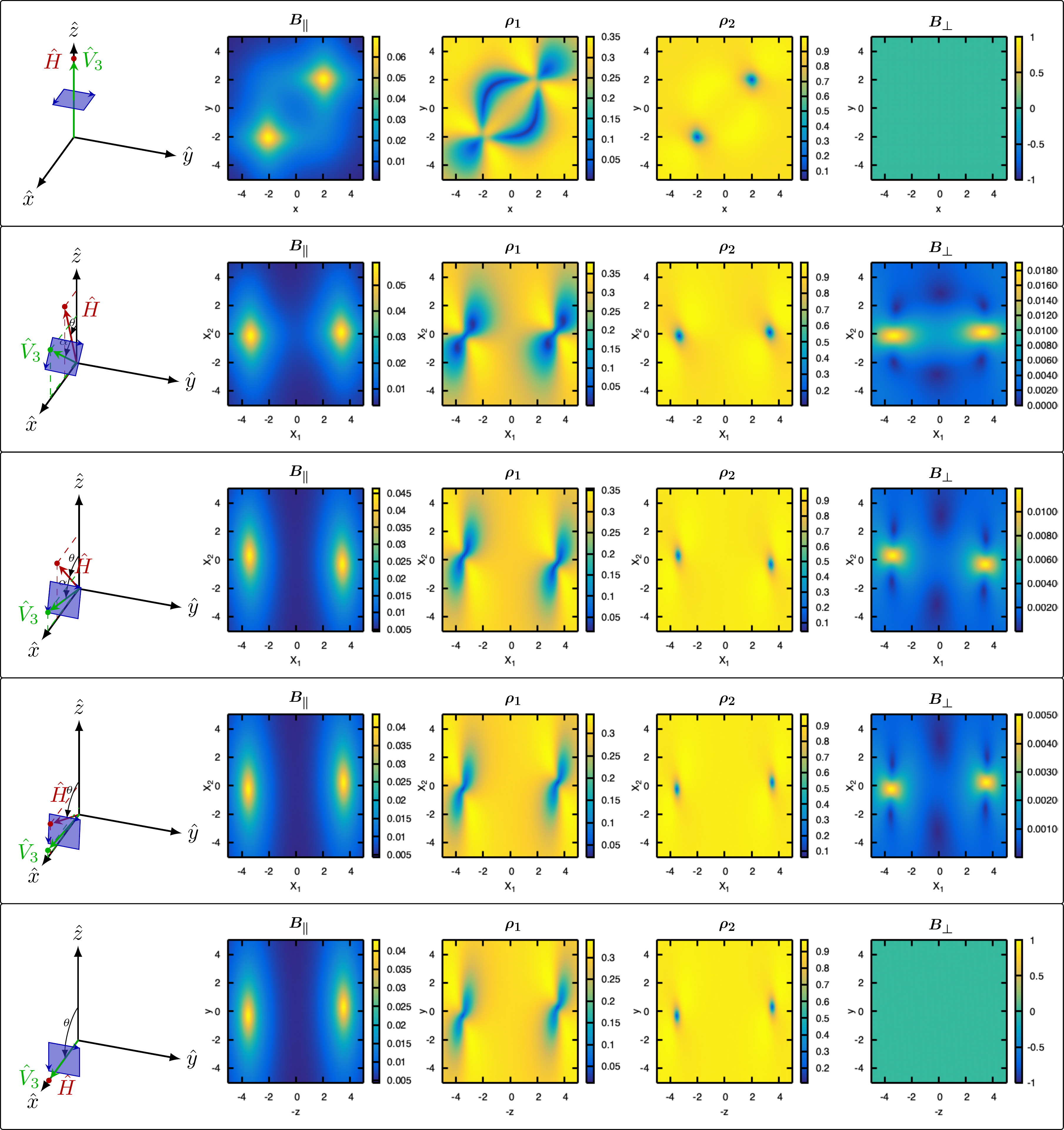}
\end{center}
\caption{Winding $N=2$ solutions  for the model in \eqref{eq:sidAni}, \eqref{eq:sidPot} 
with an applied field $|H|=H_{c_1}$ in the $x-z$ plane.  The left panel shows the applied field direction $\hat{H}$ (red arrow), the normal to the vortex plane $\hat{v}_3$ (green arrow), the angle between them $\alpha$ and the vortex plane basis $(\hat{x}_1,\hat{x}_2)$ (blue arrows), which are orthogonal to $\hat{v}_3$. We choose $\hat{x}_1$ to be the unit vector in the direction of the component of $H$ orthogonal to $v_3$, and $\hat{x}_2=\hat{v}_3\times\hat{x}_1$.
In the top and bottom rows, $\hat{v}_3=\hat{H}$, so we take $\hat{x}_1=(1,0,0)$ and $\hat{x}_1=(0,0,-1)$ respectively.   Note that $B_\parallel = B\cdot \hat{v}_3$ (along the vortex string), $\rho_\alpha$ is the magnitude of condensate $\psi_\alpha$ and $B_\perp = |B -B_\parallel \hat{v}_3|$ (orthogonal to the vortex line).}
\label{fig:sidHcvortex1}
\end{figure}

\begin{figure}
\begin{center}
\includegraphics[width = \linewidth]{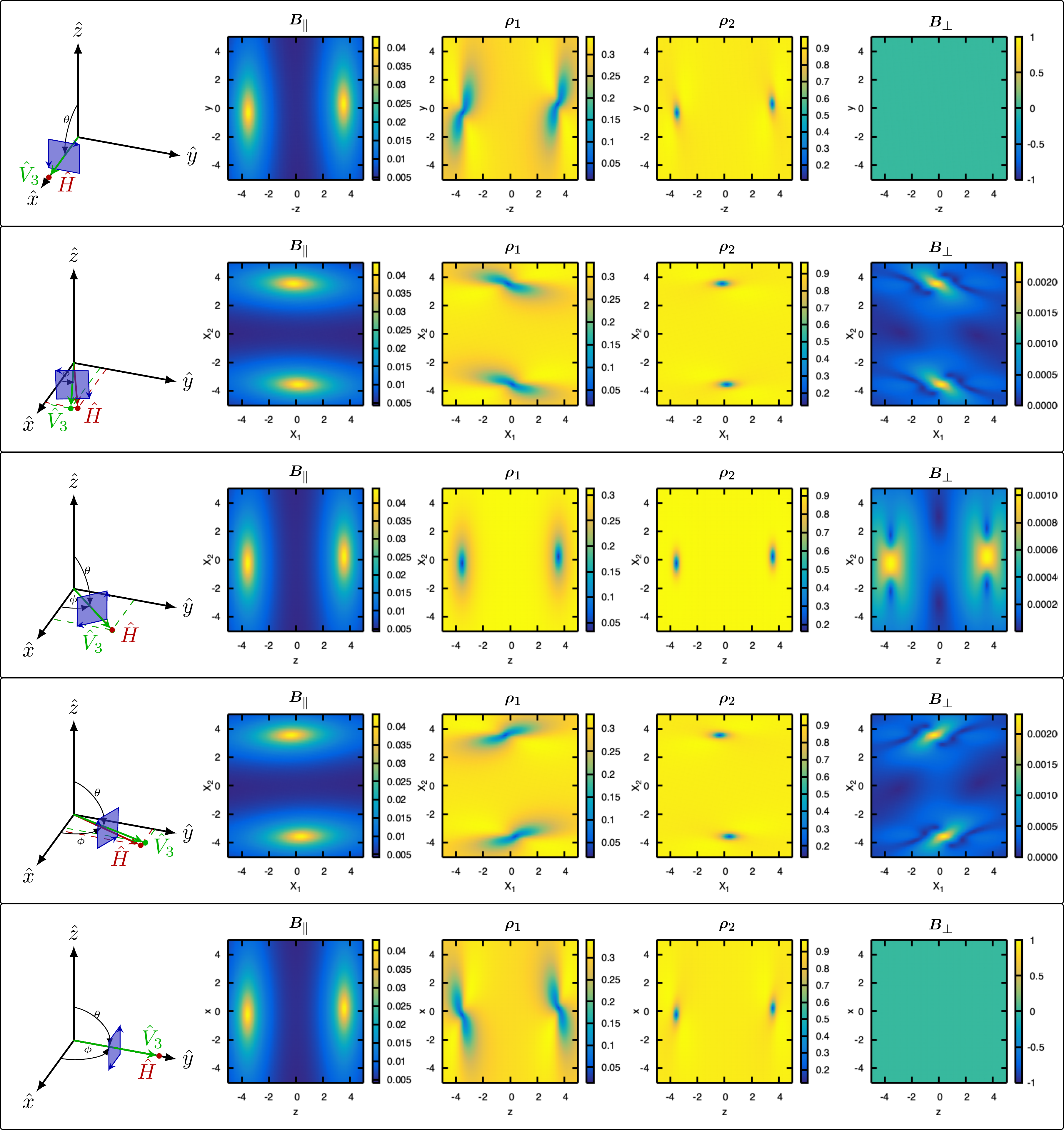}
\end{center}
\caption{Winding $N=2$ solutions for the model in \eqref{eq:sidAni},  \eqref{eq:sidPot} with applied field $|H|=H_{c_1}$  in the $x-y$ plane.  Notation and labels are the same as in \figref{fig:sidHcvortex1}. }
\label{fig:sidHcvortex2}
\end{figure}

\begin{figure}
\begin{center}
\includegraphics[width = \linewidth]{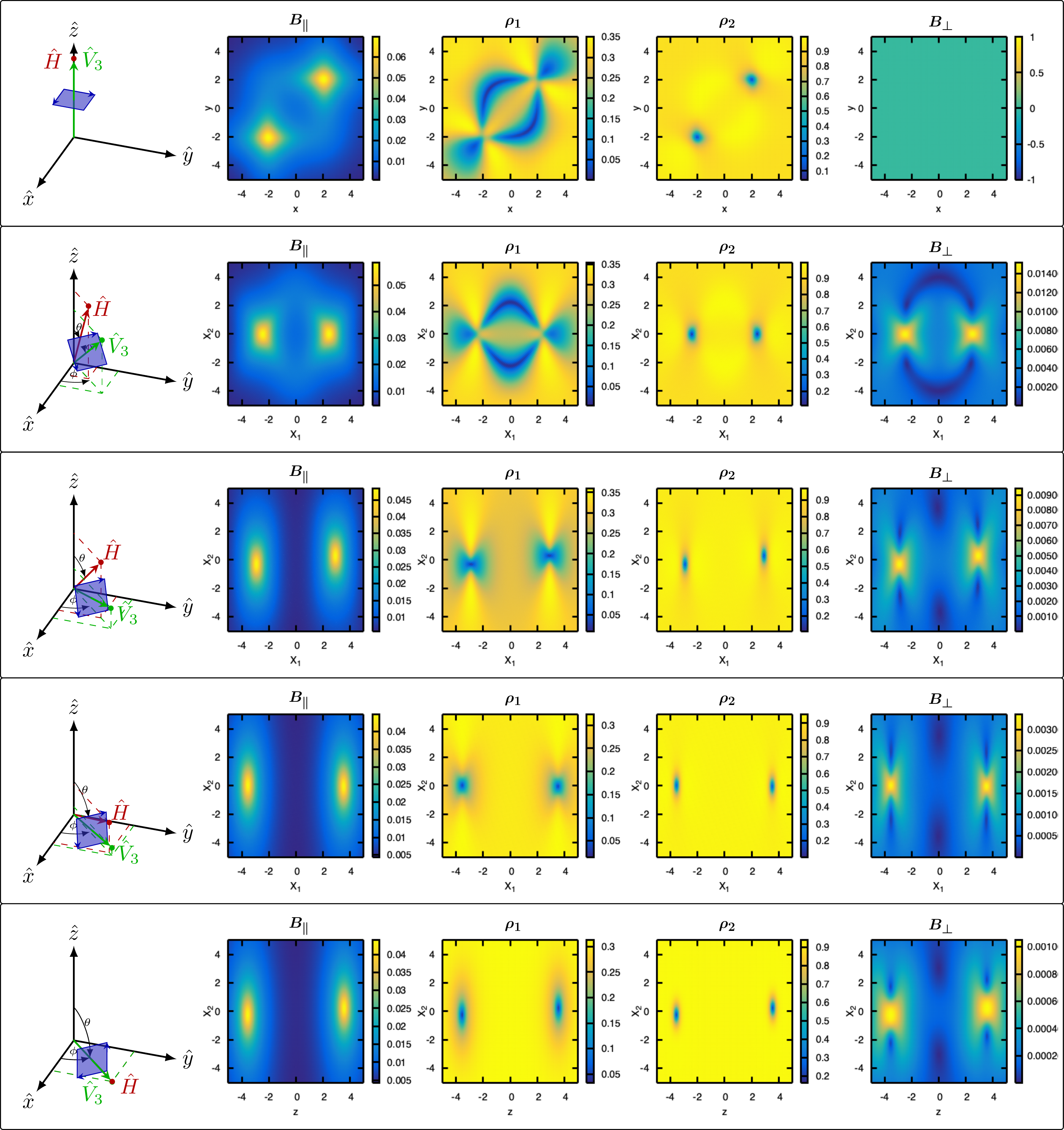}
\end{center}
\caption{Winding $N=2$ solutions for the model in \eqref{eq:sidAni},  \eqref{eq:sidPot} with applied field $|H|=H_{c_1}$  in the plane $x=y$.  Notation and labels are the same as in \figref{fig:sidHcvortex1}. }
\label{fig:sidHcvortex3}
\end{figure}

We have also plotted the $N=2$ configurations for various applied field directions that were found in the process of finding $H_{c_1}$ in \figref{fig:sidHcvortex1} for the $x-z$ plane, \figref{fig:sidHcvortex2} for the $x-y$ plane and \figref{fig:sidHcvortex3} for the $x=y$ plane. We can see that away from the crystalline axes $(\hat{x}, \hat{y}, \hat{z})$ we see substantial local magnetic field twisting, shown in the final panel of each row.  This shows that the magnitude of the magnetic field orthogonal to $v_3$ ($\sqrt{|B|^2 - B_3^2}$) is as high as $20 \%$ of $B_3$.  This will cause the magnetic field to twist direction in the plane.  This is not a surprise as $\hat{v}_3$ and $H$ are misaligned.

We also observe that the for particular orientations of applied field $H$ the vortex zeroes of the two components $\rho_1$ and $\rho_2$ are not co-centred, thus forming a so called Skyrmion \cite{garaud2013chiral}. This is a feature of a number of anisotropic models, but here it is driven by the coupled gradient terms given by $Q^{12}$.  As a result the Skyrmions only appear for the top part of the hemisphere of applied fields $H$ (when close to the $\hat{z}$ axis). This is a result of the form of $Q^{12}$ which couples the gradients in the basal plane.  Note that if we had assumed (as is usual) that $\hat{v_3}$ and $H$ were aligned we would have found a much larger region for Skyrmions, however in true physical systems it is energetically favourable for vortex lines to orient themselves closer to the basal plane (hence away from the $\hat{z}$-axis) making vortex splitting less favourable. 

\begin{figure}
\begin{center}
\includegraphics[width = 1.04\linewidth]{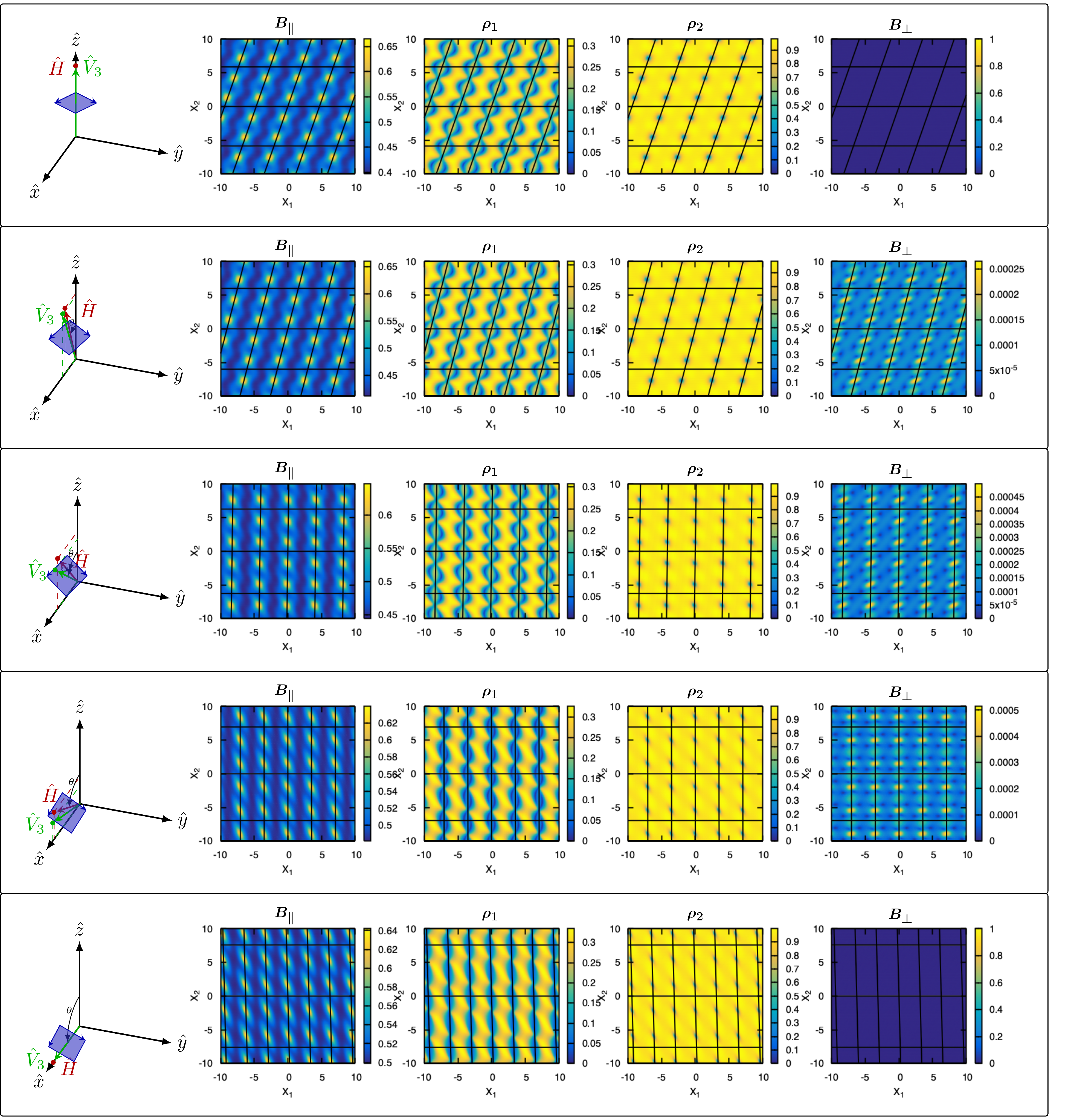}
\end{center}
\caption{Lattice solutions for the model in \eqref{eq:sidAni},  \eqref{eq:sidPot} with applied field $|H|=0.6$ in the $x-z$ plane.  The left panel shows the applied field direction $\hat{H}$ (red arrow), the normal to the vortex plane $\hat{v}_3$ (green arrow), the angle between them $\alpha$ and the vortex plane basis $(\hat{x}_1,\hat{x}_2)$ (blue arrows), which are orthogonal to $\hat{v}_3$. Note that we systematically choose $\hat{x}_1=\hat{v}_1$, parallel to one of the period vectors, and $\hat{x}_2=\hat{v}_3\times\hat{x}_1$ (which generically differs from $\hat{v}_2$). The optimal unit cell has $N=2$ and is marked in black. Note that $B_\parallel = B\cdot \hat{v}_3$ (along the vortex string), $\rho_\alpha$ is the magnitude of condensate $\psi_\alpha$ and $B_\perp = |B - B_\parallel \hat{v}_3|$ (orthogonal to the vortex line). }
\label{fig:sidLattice1}
\end{figure}

\begin{figure}
\begin{center}
\includegraphics[width = 1.1\linewidth]{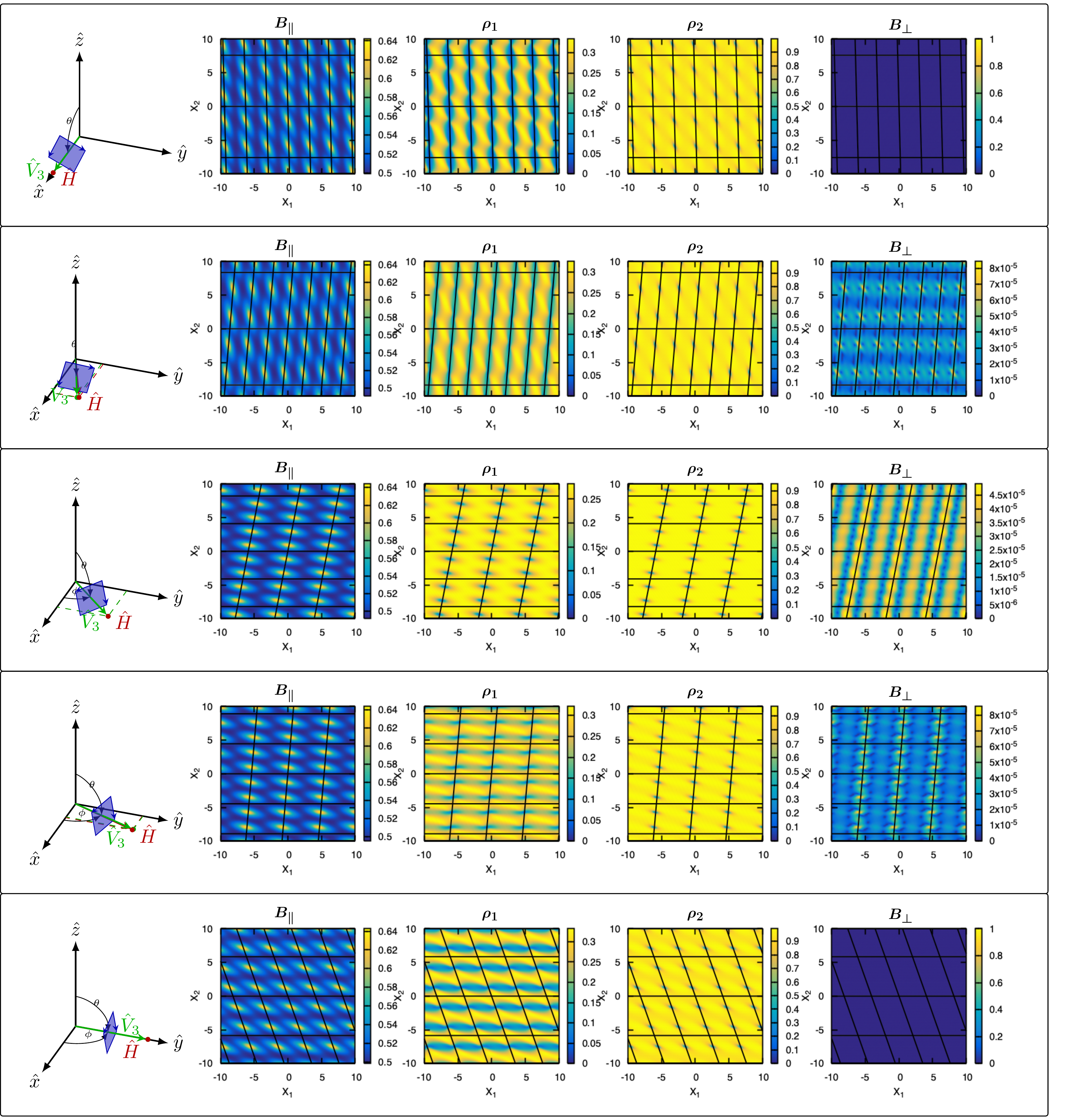}
\end{center}
\caption{Lattice solutions for the model in \eqref{eq:sidAni},  \eqref{eq:sidPot} with applied field $|H|=0.6$ in the $x-y$ plane.
Notation and labels as in \figref{fig:sidLattice1}. }
\label{fig:sidLattice2}
\end{figure}

\begin{figure}
\begin{center}
\includegraphics[width = 1.1\linewidth]{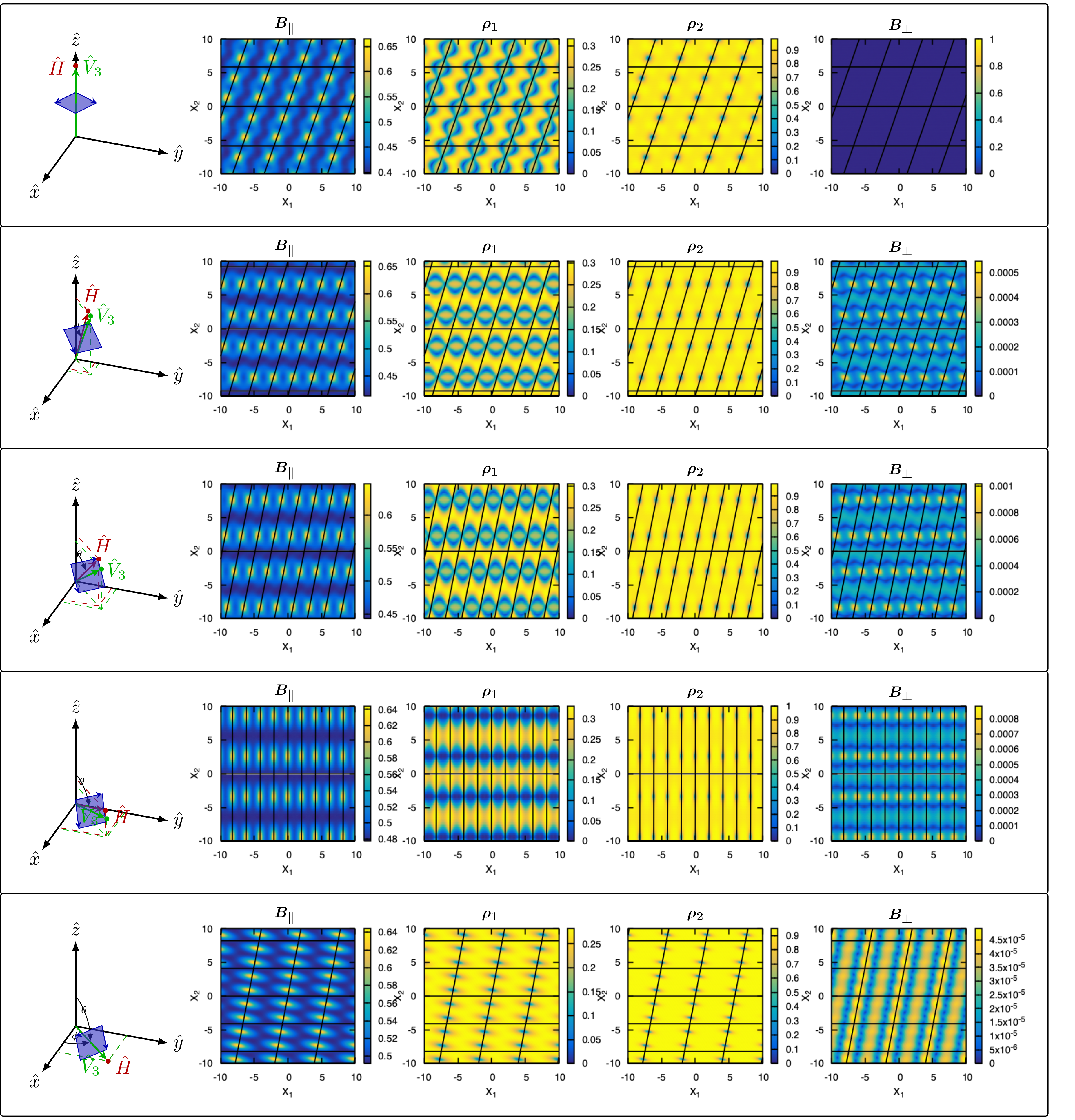}
\end{center}
\caption{Lattice solutions for the model in \eqref{eq:sidAni},  \eqref{eq:sidPot} with applied field $|H|=0.6$ in the plane $x=y$.
Notation and labels as in \figref{fig:sidLattice1}. }
\label{fig:sidLattice3}
\end{figure}

We have also shown an example of the lattice solutions for an applied field of $|H| = 0.6$ and various applied field directions \figref{fig:sidLattice1} for the $x-z$ plane, \figref{fig:sidLattice2} for the $x-y$ (basal) plane and \figref{fig:sidLattice3} for the $x=y$ plane.  We note that the stronger the applied field the more aligned the vortex line $\hat{v}_3$ is with $H$ and also the smaller the local magnetic field twisting is.  As this deviation moves the vortices away from the region where vortex splitting (Skyrmions) is observed, this leads to the surprising result that as the applied field increases this region increases.  Note that, eventually the lattice will go through a further transition near $H_{c_2}$ where the vortices are too tightly packed and cannot split \cite{speight2023symmetries}.

The above results show the importance of moving away from considering a single applied field in $\hat{z}$ direction. When one does this the possibility of $\hat{v}_3$ being misaligned from $H$ must be allowed. It not only tilts the vortex plane significantly, it directly introduces an additional anisotropic term that affects the field configurations, as can be seen by the large magnetic field twisting.  It also directly affects the regions of parameter space that various solutions exist in.

\section{Concluding remarks}\news

In this paper we have developed methods to compute energetically optimal vortex lattices, isolated vortices and the first and second critical magnetic fields in a general spatially anisotropic $n$-component Ginzburg-Landau model of superconductivity. Our methods do not assume {\em a priori} anything about the
periodicity of the vortex lattice, or the orientation of the vortex lines relative to the applied magnetic field: both of these data are determined by the energy minimization algorithm itself. We have found that even in a simple one-component model with comparatively modest anisotropy, the vortex lines of lattices at fairly low applied field (with $|H|$ only a little above $H_{c_1}$) tilt away from the direction of $H$ by as much as $40^\circ$. While $H_{c_2}$ follows the orientation dependence predicted by standard scaling arguments in the literature \cite{blatter1992scaling}, $H_{c_1}$ certainly does not.

In a more elaborate two-component $s+id$ system, recently proposed to model high $T_c$ superconductors,  we showed that the lattices exhibit fractional vortex splitting only for applied field $H$ near the $\hat{z}$-axis (away from the basal plane). We also observed that for applied field strength $|H| \approx H_{c_1}$ the direction of $v_3$ tilts away from $H$ towards the basal plane by as much as $30^\circ$.  This tilting suppresses the vortex splitting, which relies on gradient coupling terms associated with gradients parallel to the basal plane, since the line of translation symmetry is tilted towards this plane.  Hence, the region of $S^2$ consisting of applied field directions which yield vortex splitting initially grows as the applied field strength $|H|$ is increased above $H_{c_1}$, due to the decreasing angle between $v_3$ and $H$.  As 
$|H|$ is increased further, this region shrinks and eventually disappears: as $|H|$ approaches $H_{c_2}$, the lattice tends to a (possibly distorted) Abrikosov lattice, with vortex lines parallel to $H$.

It is interesting to consider how the line tilting phenomenon predicted here might be observed experimentally. We have constructed energetically optimal bulk vortex lattices, in the idealized limit of infinite sample size, using a method that rigorously excludes all surface effects (indeed our system has no boundary even in the mathematical or practical computational sense). By contrast, experimental studies of vortex lattices in tilted applied fields tend to use scanning tunneling electron microscopy to directly image the vortex cores on the surface of a (usually modestly sized) sample \cite{hess1994flux,galvis2018tilted}. It would be naive indeed to expect this vortex lattice to simply be a slice through our bulk vortex lattice along the sample surface. Boundary effects, in particular the energetics of the induced magnetic field {\em outside} the sample (the so-called stray field) are likely to exert significant effects, particularly in the regime of low applied field where our predicted tilting phenomenon is strongest. Predicting the surface vortex core distribution is a difficult mathematical challenge which is typically attempted only approximately \cite{kogan2017determining}. Theoretical studies often assume that the vortex lines are aligned with the applied field in the bulk, and bend towards the surface normal as they approach the surface. This bending is modelled by ascribing elastic properties to the vortex lines such as a shear modulus \cite{galvis2018tilted}. Hence the bulk lattice is a key input to many surface calculations, and the common assumption that bulk vortices align with the applied field is, in anisotropic systems, likely to lead to systematic errors. Our results should be relevant, therefore, to calculations of the surface distribution of vortex cores, but such calculations lie considerably beyond the scope of the present work.

To find direct evidence of line tilting in bulk vortex lattices one would need to use an experimental technnique that probes the magnetic field within the sample, such as muon spin rotation spectrosocopy \cite{sonier2000musr}. This does not directly image the vortex cores, but provides statistical information about the strength and direction of the magnetic field across the whole sample. Extracting a clean experimental signature of line tilting detectable by $\mu$SR remains a significant challenge, however.

\subsection*{Acknowledgements}

We would like to thank Egor Babaev and Alex Wormald for useful conversations. TW would like to thank the School of Mathematics at the University of Edinburgh for funding his postdoctoral position.

This work was finalized during a visit of the authors to the Jagiellonian University, Krakow, funded by EPSRC via grant 	
EP/Y033256/1.

\bibliographystyle{unsrt}
\bibliography{bibliography}

\begin{thebibliography}{10}

\bibitem{abrikosov1957magnetic}
Alexei~A Abrikosov.
\newblock On the magnetic properties of superconductors of the second group.
\newblock {\em Soviet Physics-JETP}, 5:1174--1182, 1957.

\bibitem{speight2014crystal}
JM~Speight.
\newblock Solitons on tori and soliton crystals.
\newblock {\em Communications in Mathematical Physics}, 332:355--377, 2014.

\bibitem{speight2023symmetries}
Martin Speight, Thomas Winyard, and Egor Babaev.
\newblock Symmetries, length scales, magnetic response, and skyrmion chains in
  nematic superconductors.
\newblock {\em Physical Review B}, 107(19):195204, 2023.

\bibitem{harlandleaskspeight2023}
Derek Harland, Paul Leask, and Martin Speight.
\newblock Skyrme crystals with massive pions.
\newblock {\em Journal of Mathematical Physics}, 64(10), 2023.

\bibitem{peierls1936magnetic}
R~Peierls.
\newblock Magnetic transition curves of supraconductors.
\newblock {\em Proceedings of the Royal Society of London. Series
  A-Mathematical and Physical Sciences}, 155(886):613--628, 1936.

\bibitem{london1936theorie}
von~F London.
\newblock Zur theorie magnetischer felder im supraleiter.
\newblock {\em Physica}, 3(6):450--462, 1936.

\bibitem{gor1959microscopic}
Lev~Petrovich Gor’kov.
\newblock Microscopic derivation of the ginzburg-landau equations in the theory
  of superconductivity.
\newblock {\em Sov. Phys. JETP}, 9(6):1364--1367, 1959.

\bibitem{abrikosov}
A.A. Abrikosov.
\newblock The magnetic properties of superconducting alloys.
\newblock {\em Journal of Physics and Chemistry of Solids}, 2(3):199--208,
  1957.

\bibitem{abrikosov1952}
Alexei~A Abrikosov.
\newblock {\em Dokl. Akad. Nauk SSSR}, 86:489, 1952.

\bibitem{sigrist1991phenomenological}
Manfred Sigrist and Kazuo Ueda.
\newblock Phenomenological theory of unconventional superconductivity.
\newblock {\em Reviews of Modern physics}, 63(2):239, 1991.

\bibitem{stewart2017unconventional}
GR~Stewart.
\newblock Unconventional superconductivity.
\newblock {\em Advances in Physics}, 66(2):75--196, 2017.

\bibitem{winyard2019hierarchies}
Thomas Winyard, Mihail Silaev, and Egor Babaev.
\newblock Hierarchies of length-scale based typology in anisotropic u (1)
  s-wave multiband superconductors.
\newblock {\em Physical Review B}, 99(6):064509, 2019.

\bibitem{speight2019chiral}
Martin Speight, Thomas Winyard, and Egor Babaev.
\newblock Chiral p-wave superconductors have complex coherence and magnetic
  field penetration lengths.
\newblock {\em Physical Review B}, 100(17):174514, 2019.

\bibitem{muluneh}
Habtamu~Anagaw Muluneh, Gebregziabher Kahsay, and Tamiru~Negussie Wondim.
\newblock Theoretical study of upper critical magnetic field in superconductor
  ute\_2.
\newblock {\em Indian Journal of Physics}, pages 1--11, 2023.

\bibitem{cast_of_thousands}
Y.~F. {Wang}, H.~X. {Yao}, T.~{Winyard}, Christopher {Broyles}, Shannon
  {Gould}, Q.~S. {He}, P.~H. {Zhang}, K.~Z. {Yao}, J.~J. {Zhu}, B.~K. {Xiang},
  K.~Y. {Liang}, Z.~J. {Li}, B.~R. {Chen}, Q.~Z. {Zhou}, D.~F. {Agterberg},
  E.~{Babaev}, S.~{Ran}, and Y.~H. {Wang}.
\newblock {Observation of vortex stripes in UTe$_2$}.
\newblock {\em arXiv e-prints}, page arXiv:2408.06209, August 2024.

\bibitem{nagamatsu2001superconductivity}
Jun Nagamatsu, Norimasa Nakagawa, Takahiro Muranaka, Yuji Zenitani, and Jun
  Akimitsu.
\newblock Superconductivity at 39 k in magnesium diboride.
\newblock {\em nature}, 410(6824):63--64, 2001.

\bibitem{joynt2002superconducting}
Robert Joynt and Louis Taillefer.
\newblock The superconducting phases of upt 3.
\newblock {\em Reviews of Modern Physics}, 74(1):235, 2002.

\bibitem{kamihara2008iron}
Yoichi Kamihara, Takumi Watanabe, Masahiro Hirano, and Hideo Hosono.
\newblock Iron-based layered superconductor la [o1-x f x] feas (x= 0.05- 0.12)
  with t c= 26 k.
\newblock {\em Journal of the American Chemical Society}, 130(11):3296--3297,
  2008.

\bibitem{klemm1980lower}
Richard~A Klemm and John~R Clem.
\newblock Lower critical field of an anisotropic type-ii superconductor.
\newblock {\em Physical Review B}, 21(5):1868, 1980.

\bibitem{blatter1992scaling}
G.~Blatter, V.~B. Geshkenbein, and A.~I. Larkin.
\newblock From isotropic to anisotropic superconductors: A scaling approach.
\newblock {\em Phys. Rev. Lett.}, 68:875--878, Feb 1992.

\bibitem{wang2008critical}
Zhao-Sheng Wang, Hui-Qian Luo, Cong Ren, and Hai-Hu Wen.
\newblock Upper critical field, anisotropy, and superconducting properties of
  ${\text{ba}}_{1\ensuremath{-}x}{\text{k}}_{x}{\text{fe}}_{2}{\text{as}}_{2}$
  single crystals.
\newblock {\em Phys. Rev. B}, 78:140501, Oct 2008.

\bibitem{silaev2018non}
Mihail Silaev, Thomas Winyard, and Egor Babaev.
\newblock Non-london electrodynamics in a multiband london model:
  Anisotropy-induced nonlocalities and multiple magnetic field penetration
  lengths.
\newblock {\em Physical Review B}, 97(17):174504, 2018.

\bibitem{winyard2019skyrmion}
Thomas Winyard, Mihail Silaev, and Egor Babaev.
\newblock Skyrmion formation due to unconventional magnetic modes in
  anisotropic multiband superconductors.
\newblock {\em Physical Review B}, 99(2):024501, 2019.

\bibitem{speight2023magnetic}
Martin Speight, Thomas Winyard, and Egor Babaev.
\newblock Magnetic response of nematic superconductors: skyrmion stripes and
  their signatures in muon spin relaxation experiments.
\newblock {\em Physical Review Letters}, 130(22):226002, 2023.

\bibitem{speight2020skyrmions}
Martin Speight and Thomas Winyard.
\newblock Skyrmions and spin waves in frustrated ferromagnets at low applied
  magnetic field.
\newblock {\em Physical Review B}, 101(13):134420, 2020.

\bibitem{kidszun2011critical}
Martin Kidszun, S~Haindl, T~Thersleff, J~H{\"a}nisch, A~Kauffmann, K~Iida,
  J~Freudenberger, L~Schultz, and B~Holzapfel.
\newblock Critical current scaling and anisotropy in oxypnictide
  superconductors.
\newblock {\em Physical Review Letters}, 106(13):137001, 2011.

\bibitem{tang2017anisotropic}
Zhang-Tu Tang, Yi~Liu, Jin-Ke Bao, Chuan-Ying Xi, Li~Pi, and Guang-Han Cao.
\newblock Anisotropic upper critical magnetic fields in rb2cr3as3
  superconductor.
\newblock {\em Journal of Physics: Condensed Matter}, 29(42):424002, 2017.

\bibitem{gu2018single}
Yanhong Gu, Jia-Ou Wang, Xiaoyan Ma, Huiqian Luo, Youguo Shi, and Shiliang Li.
\newblock Single-crystal growth of the iron-based superconductor la0. 34na0.
  66fe2as2.
\newblock {\em Superconductor Science and Technology}, 31(12):125008, 2018.

\bibitem{ren1995ginzburg}
Yong Ren, Ji-Hai Xu, and CS~Ting.
\newblock Ginzburg-landau equations and vortex structure of a d x 2- y 2
  superconductor.
\newblock {\em Physical review letters}, 74(18):3680, 1995.

\bibitem{wollman1993experimental}
DA~Wollman, DJ~Van~Harlingen, WC~Lee, DM~Ginsberg, and AJ~Leggett.
\newblock Experimental determination of the superconducting pairing state in
  ybco from the phase coherence of ybco-pb dc squids.
\newblock {\em Physical Review Letters}, 71(13):2134, 1993.

\bibitem{zhang2020skyrmionic}
Ling-Feng Zhang, Yan-Yan Zhang, Guo-Qiao Zha, MV~Milo{\v{s}}evi{\'c}, and
  Shi-Ping Zhou.
\newblock Skyrmionic chains and lattices in s+ i d superconductors.
\newblock {\em Physical Review B}, 101(6):064501, 2020.

\bibitem{speight2021magnetic}
Martin Speight, Thomas Winyard, Alex Wormald, and Egor Babaev.
\newblock Magnetic field behavior in s+ i s and s+ i d superconductors:
  Twisting of applied and spontaneous fields.
\newblock {\em Physical Review B}, 104(17):174515, 2021.

\bibitem{benfenati2020magnetic}
Andrea Benfenati, Mats Barkman, Thomas Winyard, Alex Wormald, Martin Speight,
  and Egor Babaev.
\newblock Magnetic signatures of domain walls in s+ i s and s+ i d
  superconductors: Observability and what that can tell us about the
  superconducting order parameter.
\newblock {\em Physical Review B}, 101(5):054507, 2020.

\bibitem{wormald2021topological}
Alex Millar~Barnes Wormald.
\newblock {\em Topological Defects in Anisotropic Multicomponent
  Superconductors}.
\newblock PhD thesis, University of Leeds, 2021.

\bibitem{garaud2013chiral}
Julien Garaud, Johan Carlstr{\"o}m, Egor Babaev, and Martin Speight.
\newblock Chiral cp 2 skyrmions in three-band superconductors.
\newblock {\em Physical Review B—Condensed Matter and Materials Physics},
  87(1):014507, 2013.

\bibitem{hess1994flux}
HF~Hess, CA~Murray, and JV~Waszczak.
\newblock Flux lattice and vortex structure in 2h-nbse 2 in inclined fields.
\newblock {\em Physical Review B}, 50(22):16528, 1994.

\bibitem{galvis2018tilted}
JA~Galvis, E~Herrera, C~Berthod, S~Vieira, I~Guillam{\'o}n, and H~Suderow.
\newblock Tilted vortex cores and superconducting gap anisotropy in 2h-nbse2.
\newblock {\em Communications Physics}, 1(1):30, 2018.

\bibitem{kogan2017determining}
Vladimir~G Kogan and JR~Kirtley.
\newblock Determining the vortex tilt relative to a superconductor surface.
\newblock {\em Physical Review B}, 96(17):174516, 2017.

\bibitem{sonier2000musr}
Jeff~E Sonier, Jess~H Brewer, and Robert~F Kiefl.
\newblock $\mu$sr studies of the vortex state in type-ii superconductors.
\newblock {\em Reviews of Modern Physics}, 72(3):769, 2000.

\end{thebibliography}

\end{document}